\documentclass[preprint]{aastex}
 
\newcommand{\apg}{\gtrsim}
\newcommand{\apl}{\lesssim}
\newcommand{\cmjj}{\mbox{${\rm cm^{-2}}$}}
\newcommand{\etal}{et al.}
\newcommand{\hI}{\mbox{H\,I}}

\newcommand{\kms}{\mbox{km\ s${^{-1}}$}}
\newcommand{\lya}{\mbox{${\rm Ly}\alpha$}}
\newcommand{\lyb}{\mbox{${\rm Ly}\beta$}}

\begin{document}
 
\lefthead{Chen \& Lanzetta}
\righthead{DAMPED \lya\ ABSORBING GALAXIES}

\slugcomment{Accepted for Publication in The Astrophysical Journal}

\title{The NATURE OF DAMPED \lya\ ABSORBING GALAXIES AT $z\le 1$--A PHOTOMETRIC
REDSHIFT SURVEY OF DAMPED \lya\ ABSORBERS$^{1}$}

\author{HSIAO-WEN CHEN\altaffilmark{2}}
\affil{Center for Space Research, Massachusetts Institute of 
Technology, Cambridge, MA 02139-4307, U.S.A. \\
hchen@space.mit.edu}

\and

\author{KENNETH M. LANZETTA}
\affil{Department of Physics and Astronomy, State University of New 
York at Stony Brook \\
Stony Brook, NY 11794--3800, U.S.A. \\
lanzetta@sbastr.ess.sunysb.edu}

\altaffiltext{1}{Based in part on observations with the NASA/ESA Hubble Space
Telescope, obtained at the Space Telescope Science Institute, which is operated
by the Association of Universities for Research in Astronomy, Inc., under NASA
contract NAS5--26555.}
\altaffiltext{2}{Hubble Fellow}

\begin{abstract}

  We study the nature of damped \lya\ absorption (DLA) systems at $z\le 1$ 
using a sample of 11 DLA galaxies, for which accurate redshift measurements are
available.  Five of the 11 systems are identified in our on-going photometric 
redshift survey of DLA galaxies, while the remaining six systems are identified
by previous groups using either spectroscopic or photometric redshift 
techniques.  Absolute $B$-band magnitude of the galaxies range from $M_{AB}(B)
=-15.3$ to $-20.3$.  Impact parameter separations of the galaxy and absorber 
pairs range from $\rho = 0.31$ to $25.4\ h^{-1}$ kpc.  We first demonstrate 
that the precision of photometric redshifts is sufficient for identifying DLA 
galaxies, because DLAs are rare and their intrinsically high column density 
implies a small impact parameter of the host galaxy to the QSO line of sight.
We then adopt this first large DLA galaxy sample to study the neutral gas cross
section of intermediate-redshift galaxies and examine the optical properties of
DLA galaxies at $z\le 1$.  The results of our study are: (1) the extent of 
neutral gas around intermediate-redshift galaxies scales with $B$-band 
luminosity as $R/R_* = [L_B/L_{B_*}]^{\beta}$ with $R_*=24-30\ h^{-1}$ kpc and
$\beta = 0.26_{-0.06}^{+0.24}$ at $N(\hI)=10^{20}$ \cmjj; (2) the observed 
incidence of the DLAs versus the $B$-band luminosity of the DLA galaxies is
consistent with models derived from adopting a known galaxy $B$-band luminosity
function and the best-fit scaling relation of the neutral gas cross section at 
$M_B - 5\log h\le -17$; (3) comparison of the observed and predicted number 
density of DLAs supports that luminous galaxies can explain most of the DLAs 
found in QSO absorption line surveys and a large contribution of dwarfs ($M_B -
5\log h\ge -17$) to the total neutral gas cross section is not necessary; (4) 
of the 11 DLAs studied, 45\% are disk dominated, 22\% are bulge dominated, 11\%
are irregular, and 22\% are in galaxy groups, indicating that galaxies that 
give rise to the DLAs span a wide range of morphological types and arise in a 
variety of galaxy environment; and (5) galaxies that contain the bulk of 
neutral gas in the universe do not appear to exhibit a substantial luminosity 
evolution between $z=0$ and $z=1$. 

\end{abstract}


\newpage

\section{INTRODUCTION}

  Damped \lya\ absorption (DLA) systems observed in the spectra of background
quasi-stellar objects (QSOs) trace neutral gas regions of \hI\ column densities
(historically defined as $N(\hI) \ge 2\times 10^{20}$ \cmjj) to redshifts as 
high as the background QSOs can be observed.  They are believed to arise in 
luminous galaxies or their progenitors at high redshift.  Optical spectroscopic
surveys for DLAs, where the absorbers are detected at redshifts ranging from 
$z=1.6$ to $z=5.0$, have demonstrated that DLAs dominate the mass density of 
neutral gas in the universe, containing roughly enough gas at $z=3.5$ to form 
the bulk of the stars in present-day galaxies (Wolfe \etal\ 1995; 
Storrie-Lombardi \& Wolfe 2000; P\'eroux \etal\ 2002).  But chemical abundance 
analyses of the DLAs yield metallicities lower than the typical solar values at
all redshifts that have been measured (e.g.\ Pettini \etal\ 1999; Prochaska \& 
Wolfe 2002; Kulkarni \& Fall 2002), suggesting that DLAs do not trace the bulk 
of star formation.  Because metallicity varies as functions of morphological 
type and galactocentric distance, it is impossible to establish connections 
between the DLAs and galaxies without first identifying the absorbing galaxies.

  Despite extensive searches, only three galaxies at high redshift (Warren 
\etal\ 2001; M{\o}ller \etal\ 2002) and four galaxies at low redshift (Rao \& 
Turnshek 1998; Lane \etal\ 1998; Miller, Knezek, \& Bregman 1999; Turnshek 
\etal\ 2001; Cohen 2001; Bowen, Tripp, \& Jenkins 2001) responsible for DLAs 
have been identified both in imaging surveys and in spectroscopic follow-up.  
A number of {\em candidate} absorbing galaxies that are likely to produce DLAs
because of their small angular separations to the QSO lines of sight in deep 
Hubble Space Telescope (HST) images have been discussed by various authors 
(Steidel \etal\ 1994; Le Brun \etal\ 1997; Warren \etal\ 2001).  The results of
the searches appear to suggest that galaxies giving rise to the high \hI\ 
column density absorbers span a wide range of morphology, from dwarfs, to 
early-type galaxies, to spirals.  The diversity in the morphology of DLA 
galaxies may explain the large scatter in the metallicity measurements, but it 
introduces some challenge to the belief that these absorbers arise in gaseous
disks and therefore trace the progenitors of present-day luminous galaxies.

  It is difficult to identify galaxies that produce DLAs based on spectroscopy 
alone due to their faintness and small impact parameter to bright, background 
QSOs, where an accurate sky determination is a challenge for the spectrum 
extraction.  It has recently become clear, however, that high-redshift galaxies
can be correctly uncovered by means of photometric redshift techniques.  For 
example, comparison of photometric and spectroscopic redshifts measured for 146
galaxies in the Hubble Deep Field has established that photometric redshifts 
are accurate to a few tenths at $z<6$ over the entire redshift range tested 
(e.g.\ Fern\'andez-Soto \etal\ 2001).  In addition, comparison of photometric 
and spectroscopic redshifts measured for 89 objects in the Hubble Deep Field 
south and Chandra Deep Field south also confirms that photometric redshifts 
estimated using ground-based photometry are both accurate and reliable to 
$\sigma_z/(1+z)=0.08$ (Chen \etal\ 2003).

  The accuracy in photometric redshift measurements is sufficient to identify 
DLA galaxies, because DLAs are rare and their intrinsic high column density 
implies a small projected distance of the host galaxy to the QSO line of sight.
Specifically, the number density of DLAs per unit redshift interval is found to
be between $n(z)\approx 0.08$ and $n(z)\approx 0.02$ (Rao \& Turnshek 2000; 
Jannuzi \etal\ 1998).  Therefore, the chances of finding two DLAs in a small 
redshift interval $\Delta z\le 1$ along a single line of sight are tiny.  
Adopting the Schechter (1976) luminosity function determined by Ellis \etal\ 
(1996) with $M_{{AB}_*}(B)=-19.6$, $\alpha=-1.4$, and $\phi_*=0.0148$, we 
expect to see on average $\le 1$ galaxy of luminosity $L \ge 0.1\ L_*$ within 
an angular radius of $5''$ to a QSO line of sight over the redshift range $0 <
z \le 1$ (Figure 1).

  Here we present the first results of a photometric redshift survey of faint 
galaxies in QSO fields with known DLAs.  When selecting the targeted DLA fields
for our survey, we have relaxed the \hI\ column density threshold to $N(\hI)\ge
10^{20}$ \cmjj\ and found 21 absorption systems at $z\le 1$ satisfying the new
column density threshold.   In all five fields that we have surveyed so far, we
have identified galaxies or galaxy groups at the same redshift as the DLAs.  
Comparison of our photometric redshifts and known spectroscopic redshifts of 11
galaxies in these fields confirms that the photometric redshift measurements 
are accurate and precise enough not only to identify the DLA galaxies, but also
to study the surrounding galaxy environment.  Combining the DLA galaxies 
identified in our survey with those known in the literature, we have collected
11 DLA galaxies at $z\le 1$ for which robust redshift measurements are 
available.  This is more than half of the total number of DLAs known at $z\le 
1$ and is the first large sample of galaxies that are known to trace the bulk 
of neutral gas in the universe.  As we shall show in this paper, the DLAs 
appear to be drawn from the same parent population as luminous field galaxies 
with $M_{{AB}_*}(B)\le -17$ ($>0.1\ L_{B_*}$) and a large contribution from
dwarf galaxies to the total \hI\ gas cross section is not necessary.

  The first part of the paper describes the strategy and observations.  We 
describe the survey design and observation strategies in \S\ 2 and observations
in \S\ 3.  Image processing and photometric redshift analysis are discussed in 
\S\ 4.  Descriptions of individual fields are presented in \S\ 5.  The second
part of the paper describes interpretations of the data.  We present the first
large sample of DLA galaxies in \S\ 6.  Results of a study to understand the 
correlation between the properties of the DLAs and the properties of the DLA 
galaxies are presented in \S\ 7, including the neutral gas cross section of 
galaxies at intermediate redshift and the morphology and intrinsic luminosity 
of the DLA galaxies.  Comparisons with previous results are presented in \S\ 8.
Finally, a summary is presented in \S\ 9.  We adopt a $\Lambda$ cosmology, 
$\Omega_{\rm M}=0.3$ and $\Omega_\Lambda = 0.7$, with a dimensionless Hubble 
constant $h = H_0/(100 \ {\rm km} \ {\rm s}^{-1}\ {\rm Mpc}^{-1})$ throughout 
the paper.

\section{A PHOTOMETRIC REDSHIFT SURVEY OF DLA GALAXIES}

  Over the past two years, we have conducted an imaging survey through multiple
bandpasses of QSO fields with known DLAs.  The goal of the survey is to 
identify a large sample of galaxies that produce DLAs at $z\le 1$ by means of 
photometric redshift techniques.  The photometric redshift techniques determine
the redshift of a galaxy by comparing the observed spectral energy distribution
(SED) established from a compilation of broadband photometric measurements with
a grid of SED templates using a maximum likelihood method.  Our survey differs 
from previous studies to look for DLA galaxies in that photometric redshift 
measurements together with well understood measurement uncertainties are 
obtained for all galaxies in the field, including those close to the QSO lines 
of sight.  Therefore we are able to not only obtain a quantitative measure of 
the confidence of the identification of each DLA galaxy, but also to study the 
clustering properties of these absorbing galaxies.

  The choice of filters and the required depths for the imaging survey were 
determined based on results from previous studies together with those of our 
own simulations in order to secure accurate photometric redshift measurements.
It is understood that photometric redshift errors arise from two different 
sources, template-mismatch variance and photometric uncertainties 
(Fern\'andez-Soto \etal\ 2001; Chen \etal\ 2003).  The former---which is due to
the finite number of templates employed in the redshift likelihood 
analysis---dominates the redshift uncertainties for bright galaxies and the 
latter dominates the redshift uncertainties for faint galaxies.  Three 
important lessons were learned from these results.  First, $U$-band photometry 
is important for accurate photometric redshift measurements for objects at $z
\le 0.4$ (Connolly \etal\ 1995; Chen \etal\ 2003).  Second, additional 
near-infrared photometry helps to distinguish between ambiguous optical 
spectral discontinuities of galaxies at low and high redshifts (Connolly 
\etal\ 1997; Fern\'andez-Soto, Lanzetta, \& Yahil 1999; Chen \etal\ 2003).
Third, photometric redshift measurements are reliable only for galaxies that 
are detected at the 30 $\sigma$ level of significance (Yahata \etal\ 2000).

  The imaging survey was therefore designed to obtain a suite of optical and 
near-infrared images of the QSO fields, the most important of which were
near-infrared $J$, $H$, and $K$ images for all DLA fields and a $U$-band image 
for the DLAs at $z\le 0.4$.  Furthermore, the required imaging depths were 
estimated from adopting a typical Scd galaxy template and assuming a 
no-evolution scenario.  We found that an $L_*$ galaxy at $z=0.8$ would have 
$I_{AB}\approx 21.8$ and therefore it was necessary to reach a 5 $\sigma$ 
limiting magnitude of $AB=26$ in the optical bandpasses for identifying 
galaxies of luminosity $L\ge 0.1\,L_*$ at this redshift, corresponding to 
$I_{AB}\approx 24.5$.  In the near-infrared bands, we aimed at a uniform 
$5\,\sigma$ limiting magnitude $AB=24$, which in the $K$ band corresponds to
$0.1\,L_*$ for a non-evolving Scd-type galaxy at $z=1$.

\section{OBSERVATIONS}

\subsection{Optical Imaging Observations}

  Optical images of the fields surrounding the QSOs AO0235$+$164, 
EX0302$-$2223, PKS0439$-$433, HE1122$-$1649, and PKS1127$-$145 were obtained 
using the direct imager with a Tek\#5 CCD on the 2.5 m du Pont telescope at the
Las Campanas Observatory. This camera has a plate scale of 0.259 arcsec 
pixel$^{-1}$ at the Cassegrain focus of the telescope, covering an $8.8'\times 
8.8'$ field of view.  The optical imaging data presented in this paper were 
taken separately in three different runs in January 2001, February 2002, and 
November 2002.  The observations were carried out in a series of five to nine 
exposures of between 300 to 900 s each and dithered by between 10 to 20 arcsec 
in space to remove deviant pixels.  Standard star fields selected from Landolt 
(1992) were observed through each of the $UBVRI$ bandpasses per night. 
Flat-field images were taken both at a white screen inside the dome and at a 
blank sky during the twilight periods every night.  Bias frames were also 
obtained every night to remove the camera bias level.

  Additional optical images of AO0235$+$164 and EX0302$-$2223 obtained with HST
using the Wide Field and Planetary Camera2 (WFPC2) with the F450W and F702W 
filters were accessed from the HST data archive.  The images of AO0235$+$164 
were taken in June 1994 with the QSO placed at the center of the Wide Field
Camera 3 in both filters.  The observations with the F450W filter were carried 
out in two exposures of 600 s each.  The observations with the F702W filter 
were carried out in two exposures of 300 s each.  The images of EX0302$-$2223 
were taken in June 1994 with the QSO placed at the center of the Planetary 
Camera in both filters.  The observations with the F450W filter were carried 
out in four exposures of between 400 and 600 s each.  The observations with the
F702W filter were carried out in six exposures of 600 s each.  A journal of the
optical imaging observations is listed in columns (1)--(5) of Table 1.

\subsection{Infrared Imaging Observations}

  Near-infrared $JHKs$ images of the six DLA fields were also obtained on the 
2.5 m du Pont telescope using the Cambridge Infrared Survey Instrument (CIRSI; 
Beckett \etal\ 1998) in October 2000 and January 2001, and the Wide Field IR
camera (WIRC; Persson \etal\ 2001) in February 2002.  These instruments contain
a sparse mosaic of four 1024$^2$ Rockwell Hawaii HgCdTe arrays and have a plate
scale of 0.196 arcsec pixel$^{-1}$ in CIRSI and 0.199 arcsec pixel$^{-1}$ in 
WIRC.  For the purpose of identifying galaxies close to the QSOs, we used only 
one array for the IR imaging of the DLA fields.  The observations were carried 
out in between five and nine sets of four exposures 45 to 75 s in duration with
dither offsets of between 10 and 20 arcsec in a rectangular pattern.  Between 
five and six IR standard stars selected from Persson \etal\ (1998) were 
observed every night.  Flat-field images were obtained at a white screen inside
the dome and were formed from differences of equal length exposures with the 
dome lamps on and off.  A journal of the near-infrared imaging observations is 
listed Table 1 as well.

\section{ANALYSIS}

\subsection{Image Processing and Galaxy Photometry}

  Individual exposures obtained with HST/WFPC2 were reduced using standard 
pipeline techniques.  Individual exposures of optical images obtained with the
Tek\#5 CCD were processed according to the following procedures.  First, bias
and overscan were subtracted from individual frames.  Next, a flat-field image
was formed for each bandpass by median filtering individual flat-field frames 
taken during twilight or inside the dome every night.  Most of the individual 
science images were flat-fielded using dome flats, except that in a few cases 
where twilight flats served better to remove hot spots in the frames.  Finally,
a fringe frame in the $I$ band was formed by median filtering the individual 
unregistered and flat-fielded $I$ frames obtained in each QSO field every night
and was subtracted from these $I$ frames after being scaled by the ratio of the
medians in the images.  Individual exposures obtained with CIRSI and WIRC were 
processed using the data reduction pipeline described in Chen \etal\ (2002).

  The processed individual images were registered to a common origin, filtered 
for deviant pixels based on a $5\,\sigma$ rejection criterion and a bad pixel 
mask formed using the flat-field frames, and stacked according to a weighting 
factor that is proportional to the inverse of the sky variance.  A summary of
the quality of the stacked images is presented in Table 1, which lists for each
field the instrument, plate scale, filter, total exposure time, full width at 
half maximum (FWHM) of the median point spread function (PSF), and the 
$5\,\sigma$ limiting magnitude in a one-arcsec diameter aperture.  In column 
(8) of Table 1, we also list for each system the expected apparent magnitudes 
in individual bandpasses of an Scd-type $L_*$ galaxy ($M_{AB_*}(B)-5\log h =
-19.6$) at the redshift of the DLA $z_{\rm DLA}$.

  Registration solutions that include rotation, linear offsets, and scaling in 
both $x$ and $y$ directions between the optical and near-infrared images were
derived based on common stars identified in each QSO field.  Because the 
near-infrared images have on average smaller PSFs, a coadded $J$, $H$, and $K$
(if available) image of each field was adopted as an image template for object 
detection using the SExtractor program (Bertin \& Arnout 1996).  The detection
threshold was set to be a group of contiguous pixels of signal-to-noise ratio 
$\apg 1.2$ with an area coverage comparable to the size of the medium PSF.  One
product of the SExtractor program is an object segmentation map of each field, 
which indicates the sizes and shapes of all the detected objects.  Instead of 
registering images of different bandpasses, the object segmentation map of each
field was registered to individual optical images according to the registration
solution.  The optical extents of individual objects were broadened by the 
difference in the optical and IR PSFs to ensure accurate flux measurements over
consistent apertures.  Object fluxes were measured by summing up all the 
photons in the corresponding apertures in the segmentation map.  Flux
uncertainties were estimated from the mean variance over the neighboring sky
pixels.

  The photometry of the space-based F450W and F702W images was calibrated using
the zero points provided by the HST.  Photometric calibrations of the 
ground-based optical $UBVRI$ images were obtained by fitting a linear function,
which included a zero-point, color, and extinction corrections, to the Landolt 
standard stars observed on the same nights.  Photometric calibrations of the 
$JHKs$ images were obtained by fitting a constant term that accounts for the 
zero-point correction to a number of the Persson standard stars observed 
throughout every night.  The RMS dispersion of the best-fit photometric 
solution was found to be 0.05 mag for the optical images and $\le$ 0.02 mag for
the near-infrared images.  Extinction corrections of individual bandpasses in 
each QSO field were estimated and applied using the Galactic extinction map 
published in Schlegel, Finkbeiner, \& Davis (1998).

\subsection{Photometric Redshifts and Uncertainties}

  We adopted the photometric redshift technique originally developed by 
Lanzetta and collaborators (Lanzetta, Yahil, \& Fern\'andez-Soto 1996; 
Fern\'andez-Soto, Lanzetta, \& Yahil 1999) but modified according to the 
descriptions in Chen \etal\ (2003) to determine galaxy redshifts based on 
ground-based photometry.  We included six galaxy spectrophotometric templates 
as presented in Fern\'andez-Soto \etal\ (1999) and Yahata \etal\ (2000) that
cover the spectral interval of $\lambda = 300-25000$ \AA\ and span a range of
spectral types---from elliptical or S0 (E/S0), Sab, Scd, irregular (Irr), to 
starburst.  These empirical templates have already included some amount of dust
reddening that is intrinsic to the host galaxies and therefore no artificial 
reddening was attempted.  Uncertainties in photometric redshifts obtained based
on ground-based galaxy photometric measurements have been studied extensively 
in Chen \etal\ (2003), where the authors showed that photometric redshifts
based on the same set of SED templates are both accurate and reliable with an 
rms residual between spectroscopic and photometric redshifts of $\sigma_z/(1+z)
= 0.08$ at $z\le 1$.

  Are photometric redshift measurements precise enough for the purpose of
identifying DLA galaxies?  To address this issue, we performed the following
exercise.  First, we calculated the corresponding impact parameter $\rho$ of 1 
arcsec angular radius versus redshift as shown in Figure 1(a).  Given that a 
typical $L_*$ galaxy in the local universe is observed to have an \hI\ extent 
at $N(\hI)= 10^{20}\,\cmjj$ of $\approx$ 22 kpc (e.g.\ Broeils \& van Woerden
1994; Cayatte \etal\ 1994), we believe that DLA galaxies reside most likely 
within $\theta=4''$ angular radius at $z=0.5$ or $\theta=3''$ at $z=1$ from the
QSO lines of sight.  

  Next, we estimated the number of random galaxies that would be seen in a 
cylinder centered at a known DLA.  The size of the cylinder is set by the 
typical neutral gaseous extent of a galaxy together with photometric redshift 
uncertainties.  If each known DLA represents a galaxy, then the probability of 
finding another galaxy by chance coincidence in this "cylinder of uncertainty"
quantifies the significance of an identification using photometric 
redshifts---a low probability of chance coincidence indicates a small 
likelihood that the true absorbing galaxy is present in the same volume but 
hidden underneath the detection limit of our survey, and therefore represents a
high confidence in the photometric redshift identifications.  Taking into 
account the fact that galaxies cluster with one another, we calculated the 
number of random galaxies that would be seen in the neighborhood of a known DLA
according to
\begin{equation}
N_{\rm gal}(\ge L_0)=n_{L_0}\,\left[1+\xi(r)\right]\, dV,
\end{equation}
where $n_{L_0}$ is the number density of galaxies of luminosity brighter than 
$L_0$, $\xi(r)\equiv (r_0/r)^\gamma$ is the real-space two-point correlation 
function of field galaxies with $r$ representing the distance in co-moving
coordinates between the DLA and surrounding galaxies, and $dV$ is the co-moving
volume element enclosed by the cylinder.  Adopting the two-point correlation 
function determined by Zehavi \etal\ (2002) based on galaxies from the Sloan 
Digital Sky Survey (which is characterized by $r_0=6.14\pm 0.18\ h^{-1}$ Mpc 
and $\gamma = 1.75\pm 0.03$) and the galaxy luminosity function of Ellis 
\etal\ (1996) (which is characterized by $\Phi_* = 0.0148_{-0.0019}^{+0.0030} 
\ h^3$ Mpc$^{-3}$ and $\alpha=1.41_{-0.07}^{+0.12}$ over the redshift interval 
$0.15 < z < 0.35$ and includes low surface brightness galaxies), we found that 
the expected overdensity of galaxies in the small volume around the known DLAs 
as a result of galaxy clustering ($n_{L_B}\xi(r)\,dV$) is negligible ($\ll 
1$\%) in comparison to the expected number of random galaxies ($n_{L_B}\,dV$).
We present the expected number of random galaxies in the neighborhood of a DLA 
versus redshift in Figure 1(b) for a combination of minimum luminosity 
considered $L_0$ and survey angular radii $\theta$.

  Figure 1(b) confirms that within a small angular radius photometric redshift 
uncertainties are relatively small to introduce little/no confusions in our 
photometric-redshift based DLA galaxy survey.  Specifically, we expect to find 
$< 0.2$ galaxies brighter than $0.1\,L_*$ within $5''$ radius from a DLA at 
$z=1$ and $< 0.06$ galaxies brighter than $0.04\,L_*$ within $2''$ radius from 
a DLA at $z=1$.  If one galaxy is identified in the ``immediate'' vicinity of a
DLA using photometric redshift techniques, the probability of finding 
additional random galaxies in the same neighborhood is vanishingly small.  In 
addition, the number density of DLAs per unit redshift interval is measured to 
be between $n(z) \approx 0.08$ and $n(z) \approx 0.02$ at $z\le 1.6$ (Rao \& 
Turnshek 2000; Jannuzi \etal\ 1998).  We would find on average no more than one
DLA per 10 lines of sight over the redshift interval $0 \le z \le 1$.  The 
possible confusion that may occur because of two closely located DLAs is also 
negligible.  We therefore conclude that photometric redshift identifications of
DLA galaxies are robust.

\section{DESCRIPTIONS OF INDIVIDUAL FIELDS}

  We have completed the multi-color imaging survey for five QSO fields with
known DLAs, and identified the DLA galaxy in every case using our photometric
redshift technique.  Properties of the DLAs and DLA galaxies identified in 
these fields are described here.

\subsection{The Field toward AO 0235$+$164}

  A radio spectrum of this BL Lac quasar ($z_{\rm em}=0.94$) exhibited a strong
21-cm absorption line at $z_{\rm abs}=0.524$ (Roberts \etal\ 1976), which was 
resolved into four distinct absorption components in a subsequent monitoring 
program (Wolfe, Brigg, \& Davis 1982).  The latter observations also showed 
that the individual components have relative flux ratios varying from a few to 
$50$\% over the course of two years.  The optical spectrum of the quasar showed
strong absorption features produced by Mg\,I, Mg\,II, Fe\,II, and Mn\,II at the
same redshift (Burbidge \etal\ 1976).  The results of the optical and radio 
observations together indicate that this absorber has a high \hI\ column 
density.  Cohen \etal\ (1999) confirmed based on a UV spectrum of the QSO 
obtained with HST/STIS that the absorber is a DLA with $N(\hI) \sim 5 \times 
10^{21}$ \cmjj.  In the following paragraphs, we shall show that this field has
previously been surveyed for the host galaxy of the QSO and the DLA galaxy by 
various authors using spectroscopic techniques.

  Figure 2 shows an image of the field around the BL Lac quasar obtained with 
the direct CCD imager on the du Pont telescope using the standard Cousins $I$ 
filter.  The image is $60''$ on a side, corresponding to $\approx 250\ h^{-1}$ 
kpc at $z = 0.5$.  The inset shows a close-up image of the immediate vicinity
around the QSO obtained with HST using WFPC2 and the F702W filter.  The light 
from the background QSO in the ground-based image has been subtracted using an 
empirical PSF determined from a median of 26 stars.  The light of the 
background QSO in the space-based image has been subtracted using a model PSF
calculated in the Tiny Tim software (Krist \& Hook 1997).  Galaxies identified 
within an angular radius of $15''$ to the background QSO are marked by their 
identification numbers (ID) in the image with the corresponding measurements 
presented in Table 2.  In columns (1)--(11) of Table 2, we list for each galaxy
the ID, Right Ascension and Declination offsets, $\Delta \alpha$ and $\Delta 
\delta$, from the QSO in arcseconds ($''$), angular distance to the QSO $\Delta
\theta$, best-fit photometric redshift $z_{\rm phot}$, spectroscopic redshift 
$z_{\rm spec}$ (if available), and $AB$ magnitudes together with associated 
uncertainties in ground-based $UBVRI$ bands.

  Four objects are identified at $\Delta \theta < 15''$ with $R<23$.  Object 
\#1 at ($+1.1''$, $0.0''$) to the QSO is well resolved in the space-based
WFPC2/F702W image but is heavily blended with the background QSO in all our 
ground-based optical images.  We were unable to obtain accurate photometric
measurements for this object using the ground-based data and we estimated that 
it has an apparent magnitude of $AB({\rm F702W})\ge 21.36$ because of the 
contaminating light from the PSF of the background QSO in the WFPC2/F702W 
image.  The remaining three objects at $\Delta \theta < 15''$ to the QSO have
$z_{\rm phot}=0.32-0.68$.  Object \#2 at $2.1''$ angular distance to the QSO is
best described by a starburst template at $z_{\rm phot}=0.52$, which coincides
with the location of the DLA.  Object \#3 at $6.5''$ angular distance to the 
QSO is best described by an E/S0 template at $z_{\rm phot}=0.32$ and object \#4
at $9.8''$ angular distance to the QSO is best described by an E/S0 template at
$z_{\rm phot}=0.68$

  This field has been surveyed by various groups to locate galaxies responsible
for the intervening absorption features.  The results showed that objects \#1
and \#2, and object \#3 at ($-5.9''$,$-2.6''$) have all been identified to have
$z_{\rm spec}=0.524$, using either narrow-band imaging (Yanny, York, \& 
Gallagher 1989) or slit spectroscopy (Cohen \etal\ 1987; Stickel, Fried, \& 
K\"uhr 1988; Burbidge \etal\ 1996; Guillemin \& Bergeron 1997).  In particular,
object \#2 is found to be a broad emission-line galaxy that shows evidence of
the presence of an active nucleus (Burbidge \etal\ 1996).  It is clear that a 
group of galaxies is responsible for the known DLA, which may explain the 
multiple absorption components observed in the 21-cm data.

  In addition, we note that two galaxies further away from the QSO line of 
sight at ($+20.2''$, $-34.7''$) and ($-30.6''$, $+28.8''$) have known redshifts
at $z_{\rm spec}=0.065$ and $z_{\rm spec}=0.28$ (Stickel \etal\ 1988; Burbidge 
\etal\ 1996).  The best-fit photometric redshifts for these two galaxies are 
respectively $z_{\rm phot}=0.1$ and 0.25, supporting that photometric redshifts
determined based on the available galaxy photometry are accurate.  While our
photometric redshifts are consistent with known spectroscopic redshifts for 
most objects in this field, the discrepancy between $z_{\rm phot}$ and $z_{\rm
spec}$ for object \#3 appears to be larger than the typical rms scatter 
$\sigma_z/(1+z)=0.08$ measured in previous studies (e.g.\ Chen \etal\ 2003).  
The unusually large discrepancy inevitably casts some doubts in the reliability
of photometric redshift techniques.  We examined in details the photometry of 
this object in order to understand the source of errors, and found that the
observed SED of the galaxy has a steep slope that cannot be fit by the regular
templates.  Stickel \etal\ (1988) presented a spectrum of this galaxy, which
shows signatures of a late-type galaxy but with a weak [O\,II]$\lambda3727$ 
emission line.  The steeper slope observed in the broad-band SED and the 
narrow-band features seen in the galaxy spectrum together suggest that the
galaxy contains a significant amount of dust.  We therefore repeated the 
photometric redshift analysis for this object by including templates that are
artificially reddened according to the formula presented in Calzetti \etal\ 
(2000) and found that the observed SED of this galaxy may be described by an
Irr template at $z=0.52$ with additional reddening of $E(B-V)=1.8$.
 
  In summary, a group of galaxies is found at the redshift of the known DLA 
$z=0.525$ toward AO0235+164.  The angular distances of these galaxies to the
background QSO range from $\Delta\theta=1.1''$ to $6.4''$, corresponding to 
$\rho=4.8-28.1\ h^{-1}$ kpc.  The proximity of these galaxies to one another 
makes it difficult to associate the DLA with any one of the galaxies 
individually, rather it is likely that each of these galaxies is a potential
DLA system as revealed in the 21-cm observations and all of them together 
produce the strong DLA feature observed in the spectrum of the background QSO.
In the subsequent neutral gas cross section analysis, we shall adopt object \#2
as the primary component for which we have accurate photometric measurements.
Object \#2 has a rest-frame $B$-band absolute magnitude $M_{AB}(B) = -20.3$ and
a corresponding projected distance to the QSO line of sight $\rho=9.4\ h^{-1}$ 
kpc at $z=0.52$.  Figure 3 shows individual images of this system through 
different bandpasses on the left and a summary of the photometric redshift 
analysis on the right.  The dimension of the image clips is $\approx 40''$ on a
side.  The top panel on the right shows the observed SED established based on 
the five photometric measurements (solid points) of the galaxy, superimposed 
with the best-fit SB template (solid line).  The bottom panel on the right 
shows the redshift likelihood function with a sharp peak at $z_{\rm phot} = 
0.52$.

\subsection{The Field toward EX 0302$-$2223}

  The ultraviolet spectrum of this QSO ($z_{\rm em}=1.409$) exhibited complex,
strong candidate \lya\ absorption features at $z_{\rm abs} = 1.0095$, 0.9874, 
and 0.9690 (Lanzetta, Wolfe, \& Turnshek 1995).  The $z_{\rm abs} = 1.0095$ 
absorber also showed strong Mg\,II and Fe\,II absorption features (Petitjean 
\& Bergeron 1990) and was confirmed to be a DLA system with $N(\hI) = (2.15 
\pm 0.35)\times 10^{20}$ \cmjj\ (Pettini \& Bowen 1997; Lanzetta 1997, 
unpublished; Boiss\'e \etal\ 1998).  The $z_{\rm abs}=0.9874$ absorber turned 
out to be a blend of C\,II feature at $z_{\rm abs}=1.3284$ and N\,I at $z_{\rm 
abs}=1.0095$ (Boiss\'e \etal\ 1998).  The $z_{\rm abs} = 0.9690$ absorber 
turned out to be a \lyb\ feature at $z_{\rm abs}=1.3284$ (Boiss\'e \etal\ 
1998).  Pettini \& Bowen (1997) showed that Zn is only a factor of two less 
abundant in the $z_{\rm abs} = 1.0095$ DLA system than in the interstellar 
medium of the Milky Way, comparable to what is expected of the Milky Way at $z
\sim 1$.  In addition, high-resolution optical spectroscopy of the QSO revealed
two distinct components 36 \kms\ apart with apparent difference in dust 
depletion, $\Delta\,[{\rm Zn}/{\rm Cr}]\approx 0.5$ (Pettini \etal\ 2000).

  Figure 4 shows the field around the QSO obtained with the Planetary Camera on
board HST using the F702W filter.  The image is $33''$ on a side, corresponding
to $185\ h^{-1}$ kpc at $z=1$.  Galaxies identified at $\Delta\theta \le 15''$
in the survey are noted by their ID in the image with the corresponding 
measurements presented in Table 3.  In columns (1)--(11) of Table 3, we list 
for each galaxy the ID, $\Delta \alpha$, $\Delta \delta$, $\Delta \theta$, 
$z_{\rm phot}$, $z_{\rm spec}$ (if available), and $AB$ magnitudes together 
with associated uncertainties in the HST F450W and F702W, and ground-based $I$,
$J$, and $H$ bands.  

  We have obtained photometric redshifts for seven objects at $\Delta \theta <
15''$ with ${\rm F702W}<25.5$.  The results of the photometric redshift 
analysis show that these galaxies have redshifts ranging from $z=0.04$ to 
$z=1.05$.  Comparison of photometric and known spectroscopic redshifts for
objects \#5, \#6, and \#7 supports that photometric redshifts determined based
on the set of photometric measurements available for galaxies in this field are
accurate.  We were, however, unable to obtain reliable photometric redshifts 
for objects \#1 and \#2.  The two objects are well resolved in the space-based
images, but are blended in the PSF of the QSO in the ground-based images.  It 
is extremely challenging to obtain robust photometric measurements for these 
objects in the $I$, $J$, and $H$ bands as a result.  In addition, objects \#3 
and \#4 are blended together with each other, although they are well resolved 
from the QSO in the ground-based images.  We attempted to obtain separate 
photometric measurements for the two galaxies based on the object extent 
determined in the F702W image.  The associated ground-based photometry were 
inevitably uncertain, resulting in large uncertainties of photometric 
redshifts.  We then determined the best-fit photometric redshift for the two 
objects combined on the basis of the summed fluxes presented in Table 3 for 
object ``3$+$4''.  The results in Table 3 indicate a consistent photometric 
redshift estimate $z_{\rm phot}=0.96$ for object \#3 with or without including
the second component \#4.  

  The results of the photometric redshift analysis indicate that objects 
\#3$+$4 and \#5 are both at the redshift of the DLA.  Object \#3$+$4 is best
described by a starburst template at $z_{\rm phot}=0.96$, while object \#5 is
best described by an Sab template at $z_{\rm phot}=1.05$.  We consider object
\#5 unlikely to be the DLA galaxy because of the large angular distance $\Delta
\theta=7.6''$ to the QSO line of sight, which corresponds to $\rho = 42.7 
\ h^{-1}$ kpc at $z = 1.0095$.  Object 3$+$4 is $3.4''$ to the QSO, 
corresponding to $\rho = 18.5\ h^{-1}$ kpc at $z=1.0095$.  We therefore 
conclude that this object is the most likely DLA galaxy with $M_{AB}(B) = 
-19.3$. 

  There are, however, two objects (\#1 and \#2) at smaller angular distances to
the QSO, which Le Brun \etal\ (1997) attributed as the DLA galaxy candidates.
Following the arguments presented in \S\ 4.2, these two objects are either at
cosmologically distinct distances away from the DLA or they are physical 
associated with object \#3$+$4.  In the second scenario, it would be difficult
to associate the known DLA with either of these galaxies individually, but it
would be more likely that the DLA is produced by a group of galaxies.  At $z =
1.0095$, object \#1 would have $\rho = 6.2\ h^{-1}$ kpc and $M_{AB}(B) = 
-18.7$, and object \#2 would have $\rho = 15.2\ h^{-1}$ kpc and $M_{AB}(B) = 
-18.8$.  If a typical $L_*$ galaxy has $M_{AB}(R)=-20.8$ and $M_{AB}(B) - 
M_{AB}(R)=0.7$ at $z\sim 1$ (Chen \etal\ 2002), then objects \#1, \#2, and 
3$+$4 would be respectively 0.27, 0.3, and $0.5\ L_*$ galaxies.  Based on
the available data, we cannot rule out either of the two scenarios.  In the 
subsequent neutral gas cross section analysis, we shall adopt object \#3$+$4 as
the DLA host galaxy for which we have a confident redshift measurement.  
Individual images of objects \#3 and \#4 together through different bandpasses 
and the results of the photometric redshift analysis are presented in Figure 5.

  We cannot rule out the possibility that the two components in object 3$+$4 
are at cosmologically distinct distances, given the available data.  It is,
however, interesting to find that if the second component (\#4) is at the same
redshift, then it appears to have relatively more intrinsic reddening according
to the optical and near-infrared color, ${\rm F702W}-H_{AB}=1.1$ (in comparison
to ${\rm F702W}-H_{AB}=-0.45$ for \#3).  This is qualitatively in line with the
dust depletion gradient found in the abundance analysis of elements Zn and Cr 
presented by Pettini \etal\ (2000).

\subsection{The Field toward PKS0439$-$433}

  The spectrum of PKS0439$-$433 ($z_{\rm em}=0.593$) obtained at ultraviolet 
wavelengths using HST/FOS exhibited a number of strong absorption features from
low-ionization transitions such as Mg\,II, Fe\,II, Si\,II, and Al\,II at 
$z_{\rm abs}=0.101$ (Petitjean \etal\ 1996).  A direct measurement of $N(\hI)$ 
using the damping wings of the \lya\ absorption feature is, however, not yet 
available.  The strong associated metal absorption lines are often a robust
indicator that the absorber has $N(\hI)>10^{20}$ \cmjj\ (e.g.\ Rao \& Turnshek
2000), which in this case is supported by an independent work reported in 
Wilkes \etal\ (1992).  These authors derived $N(\hI)\sim 1\times10^{20}$ 
\cmjj\ based on the X-ray absorption observed in the spectrum of PKS0439$-$433.
While a more precise estimate of $N(\hI)$ in this absorber awaits ultraviolet 
spectroscopy of the QSO carried out at wavelength $\lambda\sim 1350$ \AA--where
the \lya\ absorption line is located, we adopt $N(\hI)=(1.0\pm 0.8)\times 
10^{20}$ \cmjj\ for this DLA in the subsequent discussion.

  Figure 6 shows the field around the QSO obtained with the direct CCD imager 
on the du Pont telescope using the standard Cousins $I$ filter.  The image is 
$60''$ on a side, corresponding to $78\ h^{-1}$ kpc at $z=0.101$.  The light 
from the background QSO has been subtracted using an empirical PSF determined 
from a median of 10 stars in the image frame.  Galaxies identified within an 
angular radius of $15''$ to the background QSO are indicated by their ID in the
image with the corresponding measurements presented in Table 4.  In columns 
(1)--(12) of Table 4, we list for each galaxy the ID, $\Delta \alpha$ and 
$\Delta\delta$, $\Delta\theta$, $z_{\rm phot}$, $z_{\rm spec}$ (if available), 
and $AB$ magnitudes together with associated uncertainties in ground-based 
$UBVIJK$ bands.

  We have identified three objects at $\Delta \theta < 15''$ with $I<24.7$ and
$z_{\rm phot}=0-0.8$.  The results of our photometric redshift analysis showed
that object \#1 at $3.9''$ angular distance to the QSO is best described by an 
Sab template at $z_{\rm phot}=0.09$.  Objects \#2 and \#3, which are further 
away from the lines of sight, are found to be at redshifts that are very 
different from the DLA.  Object \#1 is therefore identified as the galaxy 
responsible for the DLA with a rest-frame $B$-band absolute magnitude 
$M_{AB}(B) = -19.5$.  The corresponding projected distance $\rho$ of the galaxy
to the QSO line of sight is $=5.1\ h^{-1}$ kpc.  Individual images of object 
\#1 through different bandpasses and the results of the photometric redshift 
analysis are presented in Figure 7.  

  Object \#1 is the lowest-redshift DLA galaxy identified in our survey.  It 
appears to have the typical morphology of a disk galaxy and is fairly bright.
The redshift estimate of this object has been confirmed spectroscopically by 
various authors (Petitjean \etal\ 1996; Chen, Kennicutt, \& Rauch 2003 in 
preparation).  The well resolved disk structure makes the galaxy-absorber pair
a good system for comparisons between the kinematics and chemical abundances
of the absorbers and the absorbing galaxies.

\subsection{The Field toward HE 1122$-$1649}

  The combined IUE and optical spectra of this QSO ($z_{\rm em}=2.4$) exhibited
a strong Lyman limit system with associated Mg\,II and Fe\,II absorption 
features at $z_{\rm abs} = 0.685$ (Reimers \etal\ 1995).  Subsequent 
spectroscopy using HST/FOS confirmed the presence of a DLA with $N(\hI) = 
(2.82 \pm 0.97)\times 10^{20}$ \cmjj\ at $z_{\rm abs}=0.6819$ (de la Varga 
\etal\ 2000).  In addition, a more detailed analysis of different metal 
transitions identified in a Keck HIRES spectrum indicated that this DLA has a 
metallicity of $\sim\,0.1$ solar with an abundance pattern similar to 
metal-poor Halo stars (de la Varga \etal\ 2000).  Absorption profiles of the 
low ions such as Fe\,II, Mg\,II, Mn\,II also exhibited complicated, 
multi-component structures, which follow an edge-leading signature.

  Figure 8 shows the field around the QSO obtained with the direct CCD imager 
on the du Pont telescope using the standard Cousins $I$ filter.  The image is 
$60''$ on a side, corresponding to $300\ h^{-1}$ kpc at $z=0.68$.  Galaxies 
identified within an angular radius of $15''$ to the background QSO are 
indicated by their ID in the image with the corresponding measurements 
presented in Table 5.  In columns (1)--(11) of Table 5, we list for each galaxy
the ID, $\Delta\alpha$ and $\Delta\delta$, $\Delta\theta$, $z_{\rm phot}$, 
$z_{\rm spec}$ (if available), and $AB$ magnitudes together with associated 
uncertainties in ground-based $UVIJH$ bands.

  We have identified 10 objects at $\Delta \theta < 15''$ with $I<25.7$ and
$z_{\rm phot}=0-1.24$.  The results of our photometric redshift analysis 
showed that object \#1 at $3.6''$ angular distance to the QSO is best described
by a starburst template at $z_{\rm phot}=0.69$.  Object \#6 at $10.4''$ 
angular distance to the QSO has a best-fit redshift at $z_{\rm phot}=0.63$, but
with a large redshift uncertainty.  The corresponding projected distances of 
objects \#1 and \#6 are respectively $17.7$ and 51.5 $\ h^{-1}$ kpc to the QSO 
line of sight.  In comparison to object \#1, object \#6 is 2.5 times fainter 
and three times further distant away from the line of sight.  We therefore 
conclude that object \#1 is the galaxy responsible for the DLA.  It has a 
rest-frame $B$-band absolute magnitude $M_{AB}(B) = -18.8$.  Individual images 
of object \#1 through different bandpasses and the results of the photometric 
redshift analysis are presented in Figure 9.  

  It is clear from Figure 8 that the field surrounding HE1122$-$1649 has an
overdensity of galaxies about $20''$ west of the QSO.  The results of the 
photometric redshift analysis indicate that the galaxies are at $z_{\rm phot}
\approx 0.4$.  They are therefore not associated with the DLA.  Nevertheless,
the corresponding projected distance of the galaxy group to the QSO line of
sight would be $\approx 75\ h^{-1}$ kpc at $z=0.4$.  A more detailed 
examination of metal absorption features in the QSO spectrum at the redshift of
the galaxy group would help to study the ionization state and metallicity of 
the extended gas around these galaxies.

\subsection{The Field toward PKS 1127$-$145}

  The optical spectrum of this QSO ($z_{\rm em}=1.187$) exhibited a strong 
system with associated Mg\,II and Fe\,II absorption features at $z_{\rm abs} = 
0.313$ (Bergeron \& Bossi\'e 1991).  Subsequent observations in 21-cm and 
HST/FOS spectroscopy further showed that this system is also a strong 21-cm 
absorber (Lane \etal\ 1998) and has $N(\hI) = (5.1 \pm 0.9)\times 10^{21}$ 
\cmjj\ (Rao \& Turnshek 2000).  This system resembles the one toward 
AO0235$+$164 in that the strong 21-cm absorption feature was further resolved 
into five distinct components over a velocity interval of $\approx 75$ \kms\ in
follow-up 21-cm observations and exhibited variability in the absorption line
strength over the time scale of a few days (Lane 2000; Kanekar \& Chengalur 
2001).  Finally, recent spectroscopic observations of the QSO with Chandra 
suggested that the absorbing gas of the DLA has metallicities of at most 20\% 
of the typical solar values (Bechtold \etal\ 2001).

  Figure 10 shows the field around the QSO obtained with the direct CCD imager 
on the du Pont telescope using the standard Cousins $I$ filter.  The image is 
$60''$ on a side, corresponding to $\approx 190\ h^{-1}$ kpc at $z=0.313$.  
Galaxies identified within an angular radius of $15''$ to the background QSO 
are indicated by their ID in the image with the corresponding measurements 
presented in Table 6.  In columns (1)--(12) of Table 6, we list for each galaxy
the ID, $\Delta \alpha$ and $\Delta \delta$, $\Delta \theta$, $z_{\rm phot}$, 
$z_{\rm spec}$ (if available), and $AB$ magnitudes together with associated 
uncertainties in ground-based $UBVIJH$ bands.

  We have identified eight objects at $\Delta \theta < 15''$ with $I<25.0$ and
$z_{\rm phot}=0-0.95$.  The results of our photometric redshift analysis 
showed that object \#1 at $3.8''$ angular distance to the QSO is best described
by an Scd template at $z_{\rm phot}=0.27$.  Objects \#2 also at $3.8''$ angular
distance to the QSO is best described by a starburst template at at $z_{\rm 
phot}=0.33$.  Object \#6 at $9.8''$ angular distance to the QSO is best 
described by an Sab template at at $z_{\rm phot}=0.26$.  The corresponding 
projected distances of objects \#1, \#2, and \#6 are respectively $\approx 
12.2$, 12.2, and 31.5 $\ h^{-1}$ kpc to the QSO line of sight.  In spite of a 
projected distance of about three times further away, object \#6 is more than 
15 times brighter than objects \#1 and \#2.  By way of illustration, we present
in Figure 11 individual images of object \#1 through different bandpasses and 
the results of the photometric redshift analysis for this galaxy.  

  Bergeron \& Boiss\'e (1991) obtained spectra for object \#6 and a galaxy at 
($+7.4$'',$+15.9$'') to the QSO line of sight.  They measured the redshifts of
the two galaxies to be $z=0.313$ and 0.312, respectively.  Deharveng, Buat, \&
Bergeron (1995) further estimated the oxygen abundance of object \#6 to be 
approximately 30\% of the solar value, which is in an interesting agreement 
with the abundance determined for the X-ray absorbing gas by Bechtold \etal\ 
(2001).  Lane \etal\ (1998) reported a redshift identification of object \#2 at
$z=0.312$, yielding a total of three known galaxies at the redshift of the DLA.
The results of our photometric redshift survey add one more galaxy (object \#1)
to the group of galaxies at the DLA redshift.

  In summary, a group of at least four galaxies are found at the redshift of 
the known DLA $z=0.313$ toward PKS1127$-$145.  The angular distances of these 
galaxies to the background QSO range from $\Delta\theta=3.8''$ to $17.5''$, 
corresponding to $\rho=12-56\ h^{-1}$ kpc.  The rest-frame $B$-band absolute 
magnitudes of the galaxies range from $M_{AB}(B) = -16.8$ to $-20.3$.  This is
the second DLA known to arise in a group of galaxies and have resolved 
absorption components in the 21-cm observations.  Because of the proximity of 
these galaxies to the QSO line of sight, it is difficult to separate the 
contribution of either of the galaxies to the DLA.  Despite the ambiguity 
in the contributions of the group galaxies to the observed DLA, we shall 
assign the absorber to object \#2 in the subsequent gas cross section analysis 
because it is a factor of 1.7 times brighter than object \#1 at the same 
projected distance away from the sightline.

\section{A LARGE SAMPLE OF DLA GALAXIES AT $z\le 1$}

  As an important step toward understanding the physical nature of DLA 
galaxies, we need to first identify a large sample of galaxies that give rise 
to these absorption systems observed in the spectra of background QSOs.  It 
has been, however, difficult not only because of the challenges of observing 
faint galaxies close to bright, background QSO, but also because of the small 
number of DLAs known in the literature.  Various DLA surveys carried out along
random lines of sight have yielded consistent estimates from which we expect to
find three DLAs per 10 lines of sight between $z=2.5$ and 3.5 and no more than
one DLA per 10 lines of sight between $z=0$ and 1 (e.g.\ Storrie-Lombardi \& 
Wolfe 2000; Rao \& Turnshek 2000; Ellison \etal\ 2001).  Individual DLA systems
have been reported separately from targeted surveys of QSOs that occur close 
to the centers of foreground galaxies (Miller, Knezek, \& Bregman 1999; Bowen,
Tripp, \& Jenkins 2001).  Despite extensive searches, there are only a total of
21 DLAs discovered at $z\le 1$.

  We have completed a survey of galaxies in five QSO fields with known DLAs.  
In all five cases, galaxies or galaxy groups that give rise to the DLAs are 
successfully identified using photometric redshift techniques.  As discussed 
in \S\ 4.2, the precision of photometric redshifts is sufficient for 
identifying DLA galaxies, because DLAs are rare and their intrinsic high column
density implies a small impact parameter of the host galaxy to the QSO line of 
sight.  Furthermore, we continue to examine the accuracy of our photometric 
redshift measurements in every field whenever known spectroscopic redshifts are
available.  Comparison of photometric redshifts and known spectroscopic 
redshifts for 11 galaxies in these five fields confirms that the photometric 
redshifts are accurate with an rms residual between spectroscopic and 
photometric redshifts of $\sigma_z/(1+z) = 0.07$ (Figure 12).  

  In addition to the five DLA galaxies that we presented in this paper, there 
are six DLA galaxies identified by various authors.  There are therefore 11
DLA galaxies known with robust redshift measurements, which are more than half 
of the total number of DLAs known at $z\le 1$.  We briefly summarize the
properties of these six additional DLA systems below.

  {\bf LBQS0058$+$0155} ($z_{\rm em}=1.954$): Pettini \etal\ (2000) reported 
the discovery of a DLA toward this QSO at $z_{\rm abs}=0.613$ with $N(\hI) = 
(1.2 \pm 0.5) \times 10^{20}$ \cmjj\ in an HST/FOS spectrum.  A detailed study
of chemical abundances of the absorber by these authors showed that the 
absorbing gas has near-solar metallicity.  They further reported, after a 
careful examination of an HST/WFPC2 image of the field, that a late-type disk 
galaxy $1.2''$ away from the QSO is likely to be the absorbing galaxy.  This 
object has been confirmed spectroscopically to be at $z=0.6121$ (Chen, 
Kennicutt, \& Rauch 2003, in preparation) and is therefore identified as the 
DLA galaxy at $\rho = 5.7\ h^{-1}$ kpc to the QSO line of sight.  

  {\bf Q0738$+$313} ($z_{\rm em}=0.635$): Rao \& Turnshek (1998) reported the
discovery of two DLAs toward this QSO at $z_{\rm abs}=0.0912$ and $z_{\rm abs}=
0.2212$ with $N(\hI) = (1.5 \pm 0.2) \times 10^{21}$ \cmjj\ and $N(\hI) = (7.9 
\pm 1.4) \times 10^{20}$ \cmjj, respectively.  Subsequent spectroscopic surveys
of galaxies in the field uncovered a compact galaxy at $z=0.2212$ and $5.7''$ 
away from the QSO line of sight, corresponding to $\rho=14.2\ h^{-1}$ kpc 
(Turnshek \etal\ 2001; Cohen 2001).  Cohen also reported the identification of 
a galaxy at $z=0.06$ and $31.4"$ away from the QSO line of sight that might be 
responsible for the system at $z=0.0912$.  The large redshift and angular 
separations of the galaxy and absorber pair, however, raise doubts in the 
validity of the identification.  In the subsequent analysis, we adopt the 
brightness limit $AB(K) > 17.8$ as determined by Turnshek \etal\ (2001) for the
unknown absorbing galaxy.

  {\bf B2\,0827$+$243} ($z_{\rm em}=0.939$): Rao \& Turnshek (2000) reported 
the discovery of a DLA toward this QSO at $z_{\rm abs}=0.525$ with $N(\hI) = 
(2.0 \pm 0.2) \times 10^{20}$ \cmjj.  Steidel \etal\ (2002) identified a 
highly-inclined disk galaxy at $z=0.5258$ and $5.8''$ away from the background 
QSO, corresponding to $\rho=25.4\ h^{-1}$ kpc.  Rao \etal\ (2003) presented 
$BRIK$ photometric measurements of the galaxy and found that the observed SED 
of the galaxy may be described by a galaxy at $z=0.525$ that had recently 
undergone two episodes of star-forming events.

  {\bf TON\,1480} ($z_{\rm em}=0.614$): This QSO occurs behind a nearby 
early-type galaxy NGC4203 at $z=0.0036$, around $1.9'$ away ($\rho = 5.9
\ h^{-1}$ kpc at $z=0.0036$) from the galactic center.  Miller, Knezek, \& 
Bregman (1999) obtained an ultraviolet spectrum of the QSO with HST/STIS and
reported the detection of a DLA at the redshift of the foreground galaxy with 
$N(\hI) = (2.2 \pm 1.0)\times 10^{20}$ \cmjj.  In addition, they found that 
the absorbing gas is co-rotating with the gaseous disk of the galaxy.
Accompanied with the DLA are metal absorption features produced by Mg\,II, 
Fe\,II, and Mn\,II, but the elemental abundances of these species are estimated
to be significantly below the typical solar values.

  {\bf HS1543$+$5921} ($z_{\rm em}=0.807$): This QSO occurs behind a nearby low
surface brightness (LSB) galaxy SBS1543$+$593 at $z=0.009$, only $2.4''$ away 
($\rho=0.3\ h^{-1}$ kpc at $z=0.009$) from the galactic center.  Bowen, Tripp,
\& Jenkins (2001) obtained an ultraviolet spectrum of the QSO with HST/STIS and
reported the detection of a DLA at the redshift of the foreground galaxy with 
$N(\hI) = (2.2 \pm 0.5)\times 10^{20}$ \cmjj.  Accompanied with the DLA are
strong metal absorption features produced by Si\,II, C\,II, and O\,I. 

  {\bf PKS1629$+$120} ($z_{\rm em}=1.795$):  Rao \etal\ (2003) reported the
discovery of a new DLA toward the QSO at $z_{\rm abs}=0.531$ with $N(\hI) = 
(5.0 \pm 1.0) \times 10^{20}$ \cmjj\ in recent HST/STIS observations.  These 
authors also obtained $UBRJK$ images of the field and identified a disk galaxy 
at $3.2''$ away from the background QSO.  They found that the observed SED of 
the galaxy may be described by a young galaxy at $z=0.532$ that had recently 
undergone a single burst of star formation.  Follow-up spectroscopy confirms 
that this galaxy is at $z=0.529$ (Chen, Kennicutt, \& Rauch 2003, in 
preparation) and the corresponding projected distance is $\rho = 14.1\ h^{-1}$ 
kpc

  A list of the 11 DLAs with the absorbing galaxies identified is presented in
Table 7.  In columns (1)--(9) of Table 7, we list for each system the name of
the background QSO, the QSO emission redshift $z_{\rm em}$, the redshift of the
DLA $z_{\rm abs}$, $N(\hI)$, the angular distance of the DLA galaxy to the QSO
$\Delta \theta$ in arcsec, the corresponding projected distance $\rho$, the
observed brightness in the $AB$ magnitude system, the corresponding rest-frame 
$B$-band absolute magnitude $M_{AB}(B)-5\,\log\,h$, and its visual 
morphology---whether it is bulge dominated (compact), disk dominated, or
irregular.  In addition, we have shown in \S\S\ 5.1 and 5.5 that the DLAs 
toward AO0234$+$164 and PKS1127$-$145 arise in a group of galaxies.  We 
therefore included all the identified group members in Table 7.  Finally, the 
DLA at $z_{\rm abs} = 0.0912$ toward Q0738$+$313 and the DLA at $z_{\rm abs}=
0.6563$ toward PKS1622$+$23 have been surveyed extensively for the absorbing 
galaxies by various groups (Turnshek \etal\ 2001; Cohen 2001; Steidel \etal\ 
1997) but remain unidentified.  We include these two systems at the end of 
Table 7, listing the brightness limits of the absorbing galaxies, for the 
purpose of completeness.

  The galaxies presented in Table 7 represent the first large sample of 
galaxies that are found with secure redshift measurements to give rise to 11 
known DLAs.  We note, however, that two systems toward TON\,1480 and 
HS1543$+$5921 were discovered in targeted searches of DLAs with a prior 
knowledge of the foreground absorbing galaxies.  The differences in the 
selection criteria of these two systems complicates follow-up statistical
analyses.  To facilitate our subsequent analysis, we define a ``homogeneous'' 
sample of DLA galaxies that contains nine DLA systems discovered in random 
lines of sight, and an ``inhomogeneous'' sample of DLA galaxies that includes 
the two additional systems discovered in targeted DLA searches.

  The galaxy and absorber pairs in the homogeneous sample span a redshift 
interval from $z=0.1$ to $z=1$---covering 60\% of the current universe 
age---with a median redshift $\langle z\rangle=0.5245$, and have absolute 
$B$-band magnitude ranging from $M_{AB}(B)=-17.4$ to $-20.3$.  Impact parameter
separations of the galaxy and absorber pairs range from $\rho = 5.13$ to $25.4 
\ h^{-1}$ kpc with a median of $\langle\rho\rangle=13.7\ h^{-1}$ kpc.  

\section{THE NATURE OF DAMPED \lya\ ABSORBING GALAXIES}

  An important aspect of the DLA galaxy sample is that these galaxies are 
selected uniformly on the basis of known neutral gas content ($N(\hI)\ge 
10^{20}$ \cmjj).  They represent a galaxy population that contains the bulk of
neutral gas in the universe and is complementary to galaxies selected based on
the stellar content.  In this section, we shall focus on understanding the 
nature of galaxies that produce known DLAs at $z\le 1$ and study the neutral 
gas content of intermediate-redshift galaxies using the DLA galaxy sample.

\subsection{\hI\ extent of intermediate-redshift galaxies}

  We first study the \hI\ column density distribution versus impact parameter
using the inhomogeneous DLA galaxy sample and compare the distribution with 
known \hI\ surface density profiles observed in nearby galaxies.  The goal here
is to determine the extent of neutral hydrogen gas of intermediate-redshift
galaxies selected by their association with known DLAs.  Hence all of the 
galaxies that are observed to possess extended \hI\ are included in the 
analysis presented here.

  Figure 13 shows $N(\hI)$ versus $\rho$ for the 11 galaxy and absorber pairs 
in the inhomogeneous DLA galaxy sample.  Mean \hI\ surface density profiles 
measured from 21-cm observations of nearby galaxies of different morphological
type and intrinsic luminosity are presented in different curves (Cayatte 
\etal\ 1994; Uson \& Matthews 2003).  The thick error bars in the upper-right
corner indicates both the scatter of $N(\hI)$ at the Holmberg radii of Sd-type 
galaxies observed in 21-cm data and the scatter of the \hI\ extent at $N(\hI)=
10^{20}$ \cmjj\ (Cayatte \etal\ 1994).  The two DLAs found to arise in groups 
of galaxies are marked in circles.  We use different sizes of the symbols to 
indicate the intrinsic luminosity of these galaxies in the left panel and 
redshifts in the right panel (see the figure caption).  Despite the apparent 
large scatter in the \hI\ column density distribution, we see two interesting 
features.  First, the \hI\ column density observed in DLAs is not grossly 
different from the mean \hI\ distribution of nearby galaxies, although we note 
that measurements obtained in 21-cm observations are smoothed over a finite 
beam size.  Second, while the \hI\ extent of these intermediate-redshift 
galaxies appears to be comparable to that of nearby galaxies, most DLAs tend to
lie at slightly larger radii of the absorbing galaxies for a given $N(\hI)$.

  Next, we study the distribution of $M_{AB}(B)$ versus $\rho$.  The objectives
are to determine how the \hI\ extent of the galaxies scales with intrinsic 
luminosity and to obtain a quantitative estimate of the total neutral gas cross
section of individual galaxies.  The data presented in Figure 14 exhibit an 
apparent envelope that stretches to larger $\rho$ at brighter $M_{AB}(B)$, 
suggesting that brighter galaxies have more extended \hI\ gas.  To proceed with
our analysis, we adopt a model to describe the distribution of neutral hydrogen
gas around individual galaxies.  We consider both a uniform sphere model and a 
uniform disk model, and assume a power-law relation between the \hI\ extent $R$
and the intrinsic $B$-band luminosity $L_B$ of the host galaxy,
\begin{equation}
\frac{R}{R_*} = \left( \frac{L_B}{L_{B_*}} \right)^{\beta},
\end{equation}
where $R_*$ is the characteristic \hI\ radius of an $L_*$ galaxy. 

  The disk model is more favorable for the purpose of testing the hypothesis 
that the DLAs arise in the progenitors of present-day disk galaxies.  But 
because of unknown inclination and orientation of the gaseous disks, the sphere
model allows us to obtain an empirical measure of the \hI\ extent that 
represents a lower limit to the size of the underlying disks.  Here we 
determine best-fit $R_*$ and $\beta$ for both models and compare the results
at the end.

  For a uniform sphere model, we employ a maximum likelihood analysis to 
determine $R_*$ and $\beta$.  Under this scenario, the probability of finding a
galaxy $i$ of $B$-band luminosity $L_{B_i}$ in the projected distance interval 
$\rho_i$ and $\rho_i + \Delta\rho$ that produces a DLA in the spectrum of a 
background QSO is
\begin{equation}
d\,P_i = \frac{2\,\rho_i\,\Delta\,\rho}{R^2(L_{B_i})}.
\end{equation}
Equation (3) represents a simple gas cross section argument that a random 
line-of-sight is more likely to run through gaseous clouds at larger radii 
because of larger allowed area.  The likelihood of observing $n$ galaxies 
producing $n$ DLAs at $n$ different projected distances is therefore given by 
\begin{equation}
{\cal L}=\prod_{i=1}^n\,dP_i.
\end{equation}
We find from the likelihood analysis that $R_*=24\ h^{-1}\ {\rm kpc}$ and 
$\beta=0.29$ for $M_{AB_*}(B)=-19.6$ (Ellis \etal\ 1996).  The corresponding 
2-$\sigma$ error contour is presented as the thick line in the inset of Figure 
14 with the best-fit parameters marked by the solid square.  The best-fit 
scaling relation is indicated by the straight line in Figure 14 as well.  This 
result applies over the $B$-band luminosity interval $0.1\,L_{B_*} \le L_B \le 
2.0\,L_{B_*}$ spanned by the observations.  We note that the scaling relation 
is not well determined at fainter $B$-band luminosity, because there is only 
one data point at $L_B < 0.1\,L_{B_*}$ (at $M_{AB}(B) = -15.3$ and $\rho=0.31
\ h^{-1}$ kpc) which does not place a strong constraint in the likelihood 
analysis.

  For a uniform disk model, we perform a Monte Carlo analysis to account for
the uncertainties in the inclination and orientation of the underlying gaseous 
disk.  First, we assign random inclination $\theta$ and orientation $\phi$ 
angles to describe the gaseous disk around each of the 11 DLAs, allowing that 
the probability of finding a gaseous disk of $\theta$ is proportional to $\sin
\theta$.  Next, we calculate the corresponding galactocentric distance between 
the absorber and the associated galaxy according to
\begin{equation}
r = \rho \left[ 1 + \sin^2\phi \tan^2\theta \right]^{1/2}.
\end{equation}
Next, we determine a set of best-fit $R_*$ and $\beta$ by performing the 
maximum likelihood analysis summarized in the previous paragraph with the 
probability rewritten as
\begin{equation}
d\,P_i = \frac{2\,r_i\,\Delta r}{R^2(L_{B_i})}.
\end{equation}
Finally, we repeat the procedures 10,000 times.  The results from the Monte 
Carlo analysis are shown in the inset of Figure 14.  We find that the \hI\ 
extent of individual galaxies under a uniform disk model is best characterized 
by $R_* = 30\ h^{-1}$ kpc and $\beta=0.26$ as indicated by the solid circle.  
The thin dotted line represents the 95\% uncertainty contour.  

  Comparison of the best-fit $R_*$ and $\beta$ in Figure 14 from two adopted 
models shows that while there is a large systematic uncertainty in the 
characteristic radius $R_*$ because of unknown disk alignment, the scaling 
exponent $\beta$ has consistent best-fit values based on the available data.  
Combining the results of the maximum likelihood analysis and the Monte Carlo 
simulations, we conclude that the scaling relation between galaxy $B$-band 
luminosity and \hI\ extent at $N(\hI)=10^{20}$ \cmjj\ may be described by 
equation (2) with
\begin{equation}
\beta=0.26^{+0.24}_{-0.06},
\end{equation}
and
\begin{equation}
R_*= 24 - 30 \ h^{-1}\ {\rm kpc}.
\end{equation}
The errors in equation (7) indicate the 67\% one-parameter confidence interval 
of $\beta$ for a uniform gaseous disk model.  The uncertainty interval in 
equation (8) indicates the likely $R_*$ allowed by the inhomogeneous sample
with an uncertain alignment of the underlying gaseous disk.

  Cayatte \etal\ (1994) presented \hI\ surface density profiles for 84 field 
galaxies in the local universe from 21cm observations.  The data showed that 
$R_{\rm 21cm}=(1.7\pm 0.5) R_{26.5}$, where $R_{\rm 21cm}$ is the \hI\ extent 
of these galaxies at $N(\hI)=10^{20}$ \cmjj\ and $R_{26.5}$ is the extinction 
corrected optical radius ($D_0/2$ from de Vaucouleurs \etal\ 1976).  We 
estimate that $R_{26.5}=9.0\pm 1.7\ h^{-1}$ kpc for an $L_{B_*}$ galaxy, using 
nine galaxies in this field sample for which an absolute distance measure has 
been determined by the HST Key Project to Measure the Hubble Constant (Freedman
\etal\ 2001).  Consequently, we find $R_{\rm 21cm}=15.3\pm 5.3\ h^{-1}$ kpc at 
$N(\hI) = 10^{20}$ \cmjj\ for nearby galaxies.  It appears from equation (8)
that the \hI\ extent of intermediate-redshift galaxies is marginally larger
than that of local galaxies, although with a large uncertainty.

\subsection{The incidence of DLA versus galaxy luminosity}

  The scaling relation discussed in \S\ 7.1 was derived from galaxies that are
known to produce DLAs in the spectra of background QSOs.  It is characteristic
of the \hI\ extent of the galaxies selected by their association with known 
DLAs, but it is not clear whether the scaling relation is also representative 
of those galaxies that are not included in the DLA galaxy sample.  To address 
this issue, we compare the observed incidence of the DLAs as a function of 
galaxy $B$-band luminosity with models derived from adopting a known galaxy 
luminosity function and a scaling relation between the total \hI\ gas cross 
section and galaxy $B$-band luminosity.  The objective is to determine how the 
neutral gas cross section is distributed among galaxies of different intrinsic
luminosity.  

  We restrict our analysis to the homogeneous sample, because the models 
derived from a known galaxy luminosity function is equivalent to a 
volume-limited galaxy survey (see equation 9 below).  The DLAs in the 
homogeneous sample are selected from random lines of sight and therefore 
satisfy the requirement.  This approach also allows us to obtain an independent
estimate of the scaling relation between neutral gaseous extent and galaxy 
$B$-band luminosity for comparison with the results presented in \S\ 7.1.

  We first examine the observed luminosity distribution of the nine DLA 
galaxies in the homogeneous sample.  Figure 15 shows that all of the nine DLAs 
selected from random lines of sight arise in galaxies brighter than $M_{AB}(B)
-5\,\log h = -17$ or equivalently $L_B = 0.1\,L_{B_*}$ for an $M_{{AB}_*}(B) = 
-19.6$ (Ellis \etal\ 1996), and none arises in fainter galaxies\footnote{We 
leave out the two systems toward Q0738$+$313 and PKS1622$+$23 in our analysis 
here because the absorbing galaxies can be arbitrarily bright or faint, if they
are hidden underneath the PSF of the background QSOs.}.  The observations 
indicate that luminous galaxies of $L_B > 0.1 \,L_{B_*}$ dominate the neutral 
gas cross section of the $z\le 1$ universe.

  There are, however, ambiguous cases like the DLAs toward AO0235$+$164 and
PKS1127$-$145, for which we found a group of galaxies that give rise to the 
known DLAs.  It is important to examine whether the conclusion that luminous 
galaxies dominate the neutral gas cross section would hold if we had selected
a different galaxy in the group for the two absorbers.  In the field toward
AO0235$+$164, object \#1 at $1.1''$ angular distance to the QSO line of sight
is only resolved in a space-based image and has $\approx 40$\% of the flux
observed in object \#2 which we have selected as the primary component of the
DLA.  Assuming that object \#1 has the same SED as object \#2, we work out that
it has $M_{AB}(B)-5\log h=19.3$.  In the field toward PKS1127$-$145, object \#1
has $M_{AB}(B)-5\log h=16.8$ and is at the same angular distance to the QSO 
line of sight as object \#2 which we have selected as the primary component of 
the DLA.  If we substitute these two galaxies as the primary components toward
the two lines of sight, then we have one galaxy (out of nine total) fainter
than $M_{AB}(B)-5\,\log h=-17$ and none fainter than $M_{AB}(B)-5\log h=16.5$. 
Therefore, the conclusion that luminous galaxies dominate the neutral gas cross
section remains valid.

  The observed incidence of DLAs produced by galaxies of different properties
may be modeled by adopting a known galaxy luminosity function and assuming that
every galaxy is surrounded by an \hI\ gaseous disk of radius $R(x)$.  We note 
that $R$ can be a function of different galaxy parameters $x$, such as 
intrinsic luminosity and/or morphological type, but for the present analysis we
shall focus on the case where $x\equiv L_B$.  The mean number of DLAs per unit 
redshift interval that we expect to find in galaxies of $B$-band luminosity in 
the interval $L_B$ to $L_B+d\,L_B$ is
\begin{equation}
d\,n = \frac{c}{H_0}
\frac{(1 + z)^2}
{\sqrt{\Omega_M(1+z)^3-(\Omega_M+\Omega_\Lambda-1)(1+z)^2+\Omega_\Lambda}}
\sigma(L_B)\Phi(L_B)\,d\left(\frac{L_B}{L_{B_*}}\right),
\end{equation}
where $c$ is the speed of light, $\sigma=\kappa \pi R^2(L_B)$ is the \hI\ 
absorbing gas cross section of a gaseous disk with $\kappa\equiv\cos\theta$
accounting for the effective impact area of the disk due to disk inclination, 
and $\Phi(L_B)$ is the galaxy luminosity function.

  Adopting a known galaxy luminosity function from typical faint galaxy surveys
and the power-law relation defined in equation (2) for $R(L_B)$, we can then 
determine the characteristic radius $R_*$ and scaling exponent $\beta$ using 
the homogeneous DLA galaxy sample.  For a given DLA, the probability of finding
a galaxy $i$ in the luminosity interval $L_{B_i}$ and $L_{B_i}+d\,L_B$ and the
impact parameter interval $\rho_i$ and $\rho_i+\Delta\rho$ is defined from 
equation (9) as
\begin{equation}
d\,P_i =\frac{(2\pi\,r(\rho_i)\,\Delta r)\,\kappa_i\,(L_{B_i}/L_{B_*})^{\alpha}\exp(-L_{B_i}/L_{B_*})\,d(L_B/L_{B_*})}{\langle\kappa\rangle\pi R_*^2 \int (L_B/L_{B_*})^{\alpha+2\beta}\exp(-L_B/L_{B_*})\,d(L_B/L_{B_*})},
\end{equation}
where $r$ is related to $\rho_i$ according to equation (5) for an inclined 
gaseous disk of $\theta_i$ and $\phi_i$, $\Phi(L_B)$ has been substituted with
a Schechter function characterized by $L_{B_*}$ and a faint-end slope $\alpha$,
and $\langle\kappa\rangle\,\pi R_*^2= 
\pi R_*^2\int\cos\theta\,p(\theta)\,d\,\theta$ is the mean gas cross section 
averaged over all probable inclination angles.  The mean gas cross section 
takes into account the fact that the probability density of finding a gaseous 
disk of $\theta$ is $p(\theta)=\sin\theta$.  If the inclination and orientation
of each of the gaseous disk are known, then $R_*$ and $\beta$ can be determined
using a maximum likelihood analysis in which the likelihood of observing an 
ensemble of nine galaxies responsible for nine previously known DLAs is the 
product of $d\,P_i$ over $i=1$--9.

  Because of uncertainties in the orientation of the underlying gaseous disks, 
we attempt to solve the best-fit scaling relation using a Monte Carlo method.  
As described in \S\ 7.1, we first assign each DLA galaxy a random inclination 
and orientation angle for the gaseous disk.  Then we perform the likelihood 
analysis to determine the best-fit scaling relation.  Finally, we repeat the 
process 10,000 times to find the most likely $R_*$ and $\beta$ that best 
describe the observations.

  The results are presented in Figure 16 for a Schechter luminosity function 
characterized by $M_{{AB}_*}(B)=-19.6$ and $\alpha=-1.4$ (Ellis \etal\ 1996).
We find that a scaling relation characterized by $R_*=58\,h^{-1}$ kpc and 
$\beta=0.69$ (the triangle in the figure) best describes the observed incidence
of DLAs as a function of $L_B$, if we apply the same scaling relation over the 
entire $B$-band luminosity interval.  The corresponding 95\% error contour is 
shown in the dashed line.  It is evident that the best-fit scaling relation is 
driven to large $R_*$ and steep $\beta$ due to a lack of DLA galaxies that are 
fainter than  $0.1\,L_{B_*}$ in our homogeneous sample.  This scaling relation 
naturally lends more weight on intrinsically luminous galaxies, but the 
predicted large gaseous extent $R_*=58$ kpc for an $L_*$-type galaxy is 
inconsistent with the impact parameter distribution presented in Figure 13 and 
is therefore unlikely to be correct.

  Motivated by the absence of faint galaxies responsible for the known DLAs, we
repeat the Monte Carlo analysis and impose a cut-off in the gas cross section 
for galaxies fainter than $0.1\,L_{B_*}$.  The best-fit scaling relation is 
indicated by the star in Figure 16 at $R_*=32\,h^{-1}$ kpc and $\beta = 0.37$.
The corresponding 95\% error contour is shown in the solid line.  The predicted
gaseous extent for an $L_*$-type galaxy is consistent with the impact parameter
distribution presented in Figure 13 and the scaling law agrees well with the 
optical Holmberg relation (1975).  For comparison, we also include in Figure 16
the best-fit scaling relation described in \S\ 7.1 for the inhomogeneous DLA 
galaxy sample (the cross).  We find that the best-fit scaling relation 
determined from imposing a gas cross section cut-off at faint magnitudes is 
consistent with the results presented in equations (7) and (8) .

  The predicted incidence of DLAs versus galaxy $B$-band luminosity based on 
the scaling relation characterized by $R_*=32\,h^{-1}$ kpc and $\beta = 0.37$
at $L_B\ge 0.1\,L_{B_*}$ is presented as the solid curve in Figure 15 for 
comparison with the observations.  The dashed curve shows the expectation from
extrapolating the scaling relation to fainter magnitudes.  According to the 
model, if the best-fit scaling relation obtained at $M_{AB}(B)\le -17$ applies 
to galaxies at fainter magnitudes, then they do not contribute to more than 
44\% of the total \hI\ gas cross section at $z<1$.  We note that of all the 21 
DLAs known at $z\le 1$, nine have been identified with galaxies of $L_B\ge 
0.1\,L_{B_*}$ and two are uncertain.  If the scaling relation applies to 
galaxies of $L_B\le 0.1\,L_{B_*}$, then it implies that almost all of the 
remaining 10 DLAs would be produced by galaxies fainter than $M_{AB}(B)=-17$. 
Because the DLA galaxies have been selected from random lines of sight by 
various authors, we find it unlikely that our DLA galaxy sample is biased 
toward luminous galaxies.  

  Furthermore, comparison of the observed DLA number density at $z<1$ and the
prediction from equation (9) offers an additional constraint on the 
contribution from faint galaxies to the total \hI\ gas cross section.  Rao \& 
Turnshek (2000) measured the number density of DLAs per unit redshift per line 
of sight and reported $n(z)=0.08_{-0.04}^{+0.06}$ at $z=0.49$.  Integrating  
equation (9) over the luminosity interval of $L_B\ge 0.1\,L_{B_*}$ for a 
luminosity function characterized by $M_{{AB}_*}(B)=-19.6$, $\alpha=-1.4$, and
$\phi_*=0.0148$ (Ellis \etal\ 1996) and a scaling relation characterized by
$R_*=32\,h^{-1}$ kpc and $\beta = 0.37$, we find that $n(z)=0.16$ at $z=0.5$. 
The predicted number density from luminous galaxies alone can already explain 
all the observed DLAs in QSO absorption line surveys, supporting that the 
scaling relation for dwarf galaxies may be substantially different from 
equations (7) and (8) and a large contribution from dwarf galaxies to the 
neutral gas cross section is not necessary. 

  In summary, our analysis shows that the observed incidence of DLAs as a 
function of galaxy $B$-band luminosity follows the expectation derived from a 
gas cross section weighted galaxy luminosity function at $M_{{AB}_*}(B)\le -17$
and that dwarf galaxies have a negligible contribution to the total \hI\ gas 
cross section.  On the other hand, a targeted DLA survey toward the low surface
brightness galaxy (HS1543$+$5921) of $M_{{AB}_*}(B))-5\log h=-15.3$ did reveal
a DLA feature in the background QSO at $\rho=0.31\ h^{-1}$ kpc.  It is 
therefore clear that dwarf galaxies ($M_{{AB}_*}(B)\ge -17$) do contain \hI\ 
gas.  Consequently, the imposed cut-off at $M_{{AB}_*}(B)=-17$ necessary to 
obtain a reasonable best-fit $R(L_B)$ suggests that dwarf galaxies have neutral
gas cross section that is scaled down from that of more luminous galaxies.

\subsection{Evolution with redshift}

  The homogeneous sample of DLA galaxies also allows us to examine the redshift
evolution of intrinsic $B$-band luminosity of the galaxies selected from known 
DLAs.  Because these DLA galaxies trace the bulk of neutral gas in the 
universe, the objective is to study whether there exists a steep evolution in
their intrinsic luminosity as reported in various magnitude-limited surveys of
field galaxies (see e.g.\ Lilly \etal\ 1996).  Figure 17 shows the distribution
of rest-frame $B$-band absolute magnitude versus redshift for the 11 DLAs 
(closed circles) that have been surveyed so far, including the two systems for
which no absorbing galaxies have yet been found (indicated by arrows).  The two
systems that are found to arise in groups of galaxies are marked with open
triangles.  For comparison, we also include in the plot the two additional 
systems from the inhomogeneous DLA galaxy sample (open circles).  Considering 
only the systems that are associated with individual galaxies (excluding the
open triangles), we find no indication of steep brightening in the intrinsic 
luminosity of the host galaxies from $z\sim 0$ to $z\sim 1$.  We note, however,
that this result is still limited by the small number of DLAs known at low 
redshift.

\subsection{Morphology and galaxy environment}

  One of the primary objectives of our study is to resolve the puzzle of the 
apparent low metallicity measured in these systems.  It is not clear whether 
the low metallicity determined for the DLAs occurred as a result of metallicity
gradient commonly observed in nearby disk galaxies (e.g.\ Henry \& Worthey 
1999; Kennicutt, Bresolin, \& Garnett 2003) or as a result of the diversity in 
the morphology of the DLA galaxies.  Based on the homogeneous sample of DLA 
galaxies, we find that the morphological types of the DLA galaxies are directly
observed to range from compact galaxies (e.g.\ the DLA toward HE1122$-$1649) 
through normal spiral galaxies (e.g.\ the DLA toward PKS0439$-$433) through low
surface brightness galaxies (e.g.\ the DLA toward EX0302$-$2223).  
Specifically, of the 11 DLAs studied, four arise in disk-dominated galaxies
(45\%), two in bulge-dominated galaxies (22\%), one in an irregular galaxy 
(11\%), and two are in galaxy groups (22\%). 

  It is evident from the available data that DLAs probe galaxies of different 
morphology and galactic environment, consistent with previous findings (e.g.\ 
Le Brun \etal\ 1997).  However, we caution possible systematic biases against 
detections of low surface brightness stellar disks in the ground-based images.
In addition, various studies have shown that galaxies exhibit different 
morphologies in different bandpasses, but we have not accounted for this 
morphological $k$-correction in our discussion.  The bias is likely to be more 
significant for galaxies at higher redshifts.  For example, the DLA galaxy 
toward EX0302$-$2223 appears to show irregular structures in the F702W 
bandpass.  At $z=1$, the observed F702W bandpass corresponds to the rest-frame
$U$-band, in which local disk galaxies often appear to be irregular (e.g.\ 
Windhorst \etal\ 2002).  High spatial resolution images obtained in the near
infrared will be needed for a more reliable assessment for this particular 
system.  For the DLA galaxies at lower redshift, future HST images obtained
through two bandpasses will be crucial for confirming our initial 
classifications using ground-based images.

  We also note that the two strongest DLAs presumed to arise in galaxy groups 
with $N(\hI)>10^{21}$ \cmjj\ exhibit unusual properties that are strikingly 
similar (Figure 13; points in open circle).  First, they exhibit variability in
the absorption line strengths over the course of a few days to six months 
(Wolfe \etal\ 1982; Kanekar \& Chengalur 2001).  Second, there exist $4-5$ 
distinct components in the 21-cm absorption profiles observed against the 
background QSOs, which span a velocity range of $\approx 70$ \kms\ (Wolfe 
\etal\ 1982; Lane 2000; Kanekar \& Chengalur 2001).  Both sets of observations 
indicate that neither of the DLAs is likely to arise in a single large \hI\ 
cloud.  These properties together suggest that the underlying gaseous 
environment is different from those in typical galaxies.  Two additional DLAs 
at $z\ge 1$ are estimated to have $N(\hI) > 10^{21}$ \cmjj, one at $z = 0.685$ 
toward Q0218$+$357 and the other at $z = 0.394$ toward Q0248$+$430 (Rao \& 
Turnshek 2000).  Future high spectral resolution 21-cm monitoring of these
absorbers will provide further clues on whether or not these very high 
column density DLAs arise preferentially in denser galaxy environment.

\section{COMPARISON WITH PREVIOUS WORK}

  The long-standing belief that the DLAs probe the gaseous progenitors of
present-day luminous galaxies at high redshift (see Wolfe 1995) is challenged 
by various reports that dwarf galaxies dominate the galaxy population that 
produces the DLAs at $z<1$ (Le Brun \etal\ 1997; Colbert \& Malkan 2002; Rao 
\etal\ 2003).  Our results have, however, demonstrated that dwarf galaxies do 
not dominate the neutral gas cross section, consistent with the results of a 
recent study of \hI\ gas cross section of nearby galaxies (Ryan-Weber, Webster,
\& Staveley-Smith 2003).  Here we compare our study with those published by 
previous authors in order to understand this discrepancy.

  Le Brun \etal\ (1997) observed six QSO fields with known DLAs at $z\le 1$
using HST/WFPC2 and identified {\em candidate} DLA galaxies based on their 
close proximity to the QSO lines of sight.  Adopting the F702W-band photometric
measurements of these candidate galaxies presented by the authors and the 
$\Lambda$ cosmology, we find that these candidates span a $B$-band absolute 
magnitude range from $M_{AB}(B)=-16.8$ to $-20.0$ with one fainter than 
$M_{AB}(B)=-17$, two in $-18 \le M_{AB}(B) \le -17$, and three brighter than 
$M_{AB}(B)=-19$.  This is {\em not} inconsistent with the predicted incidence 
of DLAs that we presented in \S\ 7.2.


  Rao \etal\ (2003) presented results from a photometric redshift survey of 
four QSO fields with known DLAs.  These authors successfully identified the 
absorbing galaxies for three systems using photometric redshift techniques and 
reported the detection of a candidate galaxy in a $K$-band image.  Rao \etal\ 
collected a sample of 14 {\em candidate} DLA galaxies at $0.05\le z\le 1$, for
which only five galaxies have secure redshifts available.  They conclude that 
low-luminosity dwarf galaxies with small impact parameters dominate the bulk of
the neutral hydrogen gas in the universe at $z\approx 0.5$ and that the highest
column density systems arise in low surface brightness dwarf galaxies.  These 
conclusions are in complete disagreement with ours presented in \S\ 7.  Here we
investigate possible reasons that may have resulted in the gross discrepancy.

  Our DLA sample overlaps with the Rao \etal\ sample in six fields, 
AO0235$+$164, EX0302$-$2223, Q0738$+$313 (the DLA at $z=0.2212$), 
B2\,0827$+$243, PKS1127$-$145, and PKS1629$+$120.  These authors attributed the
DLAs toward AO0235$+$164 and PKS1127$-$145 to the faintest galaxies in the 
groups found at the absorber redshifts.  They also attribute the DLA toward 
EX0302$-$2223 to the candidate galaxy that has been found at the closest 
angular distance to the QSO line of sight, following Le Brun \etal\ (1997).
Of the six overlapping fields, Rao \etal\ concluded that three DLAs arise in
galaxies fainter than $0.25\,L_*$, while the other three DLAs arise in galaxies
brighter than $0.5\,L_*$.  This is qualitatively consistent with the 
expectation derived in \S\ 7.2.  It appears that the large discrepancy comes 
primarily from the remaining eight DLA fields, for which no reliable redshift 
measurements are available. 
 
  However, the fields toward AO0235$+$164 and PKS1127$-$145 are known to have 
groups of galaxies at close distances to the DLAs.  Each of these galaxies is 
likely to contribute to the observed DLA features, as suggested by the 21-cm 
observations.  Consequently, the conclusion that the highest column density 
systems arise in low surface brightness dwarf galaxies needs to be 
re-evaluated.

  In addition, there is a systematic uncertainty in the definition of dwarf 
systems induced by the differences in optical and near-infrared galaxy 
luminosity, which is clearly present in Rosenberg \& Schneider (2003).  These 
authors estimated the contribution of \hI-rich galaxies to the DLA population 
at $z=0$ using an \hI\ mass function derived from a blind 21-cm survey.  While 
these authors concluded that galaxies of \hI\ masses $\approx 10^9\,M_\odot$ 
dominate the total neutral hydrogen gas, they also showed a flat distribution 
in the $N_{\rm DLA}$ versus $J$-band luminosity $L_J$ plot.  It indicates that
a large fraction of DLAs should arise in near-infrared faint galaxies.  We
emphasize, however, that whether or not ``dwarf'' galaxies have a substantial 
contribution to the neutral gas cross section is still uncertain, because a 
large scatter between \hI\ mass $M_{\hI}$ and $L_J$ is evident (see their 
Figure 4).  The corresponding $L_J$ spans at least three orders of magnitude 
for a given $M_{\hI}$, which is likely to flatten the distribution of $N_{\rm 
DLA}$ versus $L_J$. 

  In summary, we find that the luminosity distribution of the candidate DLA
galaxies in Le Brun \etal\ is consistent with the results of our analysis in
\S\ 7.2 that dwarf galaxies ($M_{AB}(B)>=-17$) do not dominate the neutral gas
cross section in the $z\le 1$ universe .  The main conclusion from Rao \etal\ 
that dwarf galaxies have significant contributions to the neutral gas cross 
section needs to be re-evaluated, when luminosity uncertainties of their DLA 
galaxies are taken into account.

\section{SUMMARY AND CONCLUSIONS}

  We presented results from a photometric redshift survey of galaxies in five 
QSO fields with known DLAs.  In all five cases using photometric redshift
techniques, we have identified the galaxies or galaxy groups that are 
responsible for the known DLAs.  We demonstrate that the precision of 
photometric redshifts is sufficient for identifying DLA galaxies, because DLAs 
are rare and because their intrinsic high column density implies a small impact
parameter of the host galaxy to the QSO line of sight.  Comparison of 
photometric redshifts and known spectroscopic redshifts for 11 galaxies in 
these fields also confirms that the photometric redshifts are accurate to 
within an rms uncertainty of $\sigma_z/(1+z)=0.07$.  Combining the results of 
our survey with known DLA galaxies that have been previously identified using 
spectroscopic or photometric redshift techniques, we have collected 11 DLA 
galaxies at $z\le 1$ for which reliable redshift measurements are available.  
This is more than half of the total number of DLAs known at $z\le 1$.  

  However, the differences in the selection criteria of these DLA systems make
any statistical analysis using the entire sample complex and difficult.  We 
have therefore defined a ``homogeneous'' sample of DLA galaxies that contains 
nine DLA systems discovered in random absorption lines surveys, and an 
``inhomogeneous'' sample of DLA galaxies that includes the two systems 
discovered in targeted DLA searches.  Redshifts of the DLA galaxies in the 
homogeneous sample range from $z=0.1$ to $z=1$---covering 60\% of the current 
universe age.  Absolute $B$-band magnitude of the galaxies range from 
$M_{AB}(B)=-17.4$ to $-20.3$.  Impact parameter separations of the galaxy and 
absorber pairs range from $\rho = 5.1$ to $25.4 \ h^{-1}$ kpc.  We determined 
neutral gas extent of luminous galaxies using the inhomogeneous DLA galaxy
sample and examined the luminosity distribution and galaxy environment of DLA 
galaxies using the homogeneous sample.  The results are summarized as follows:
  
  1. We find based on a uniform gaseous disk model that the neutral gas cross 
section of individual galaxies scales with $B$-band luminosity as $\sigma(\hI)=
\pi R^2(L_B/L_{B_*}) = \pi R_*^2 [L_B/L_{B_*}]^{2\beta}$ with $\beta =
0.26_{-0.06}^{+0.24}$ and $R_*= 24-30\ h^{-1}$ kpc at $N(\hI)= 10^{20}$ \cmjj.
The result suggests that the neutral gaseous extent of intermediate-redshift 
galaxies is marginally larger than what is observed in local galaxies.

  2. The observed incidence of the DLA galaxies as a function of galaxy 
$B$-band absolute magnitude indicates that luminous galaxies ($M_B - 5\log h\le
 -17$) dominate the neutral gas cross section in the $z\le 1$ universe.  In 
addition, the observations agree well with predictions from adopting a known 
galaxy luminosity function and the best-fit scaling relation of the total gas 
cross section at $M_B - 5\log h\le -17$.  The necessary cut-off at 
$M_{{AB}_*}(B)=-17$ suggests that dwarf galaxies possess \hI\ gas cross section
that is scaled down from luminous galaxies

  3. Comparison of the observed number density of DLAs and predictions derived 
from our model indicates that luminous galaxies ($M_B - 5\log h\le -17$) can 
already explain most of the DLAs found in QSO absorption line surveys and a
large contribution of dwarfs to the total neutral gas cross section is not
necessary

  4. Morphological types of DLA galaxies are directly observed to range from 
bulge-dominated (compact) galaxies through normal disk galaxies through low 
surface brightness galaxies, indicating that the DLA galaxy sample is roughly 
representative of the galaxy population over a large fraction of the Hubble 
time.

  5. The DLAs probe a variety of galaxy environment.  In particular, two of the
highest column density systems ($N(\hI) > 10^{21}$ \cmjj) at $z\le 1$ that
show distinct absorption components in 21-cm observations are found to arise in
groups of galaxies.

  6. Galaxies selected based on $N(\hI) \ge 10^{20}$ \cmjj\ do not appear to 
have a substantial luminosity evolution between $z=0$ and $z=1$.

  In summary, we conclude that the underlying galaxy population traced by the 
DLAs at $z\apl 1$ may be well characterized by a gas cross-section weighted 
galaxy luminosity function at $M_{AB}(B)-5\log h\le -17$.  Dwarf galaxies 
($M_{AB}(B)-5\log h\ge -17$) do not dominate the total \hI\ gas cross section, 
although they may have some contribution.

\acknowledgments We appreciate the expert assistance from the staff of the Las
Campanas Observatory.  We thank Scott Burles, Rob Kennicutt, Lynn Matthews, and
Jason Prochaska for helpful discussions.  In addition, we thank Alberto 
Fern\'andez-Soto, Miguel Roth, and Paul Schechter for helpful comments on an 
earlier draft of this manuscript.  This research was supported in part by NASA 
through the American Astronomical Society's Small Research Grant Program.  
H.-W.C. acknowledges support by NASA through a Hubble Fellowship grant 
HF-01147.01A from the Space Telescope Science Institute, which is operated by 
the Association of Universities for Research in Astronomy, Incorporated, under 
NASA contract NAS5-26555.

\newpage


\begin{tiny}
\begin{center}
\begin{tabular}{p{1.75 in}ccccccc}
\multicolumn{8}{c}{Table 1} \\
\multicolumn{8}{c}{Summary of the Optical and Near-infrared Imaging Data} \\
\hline
\hline
 & & Plate Scale &  & Exposure & FWHM & & \\
\multicolumn{1}{c}{Field} & \multicolumn{1}{c}{Instrument} & (arcsec$/$pixel) 
& \multicolumn{1}{c}{Filter} & Time (s) & (arcsec) & $AB_{5\,\sigma}$ 
& $AB_*(z_{\rm DLA})$$^a$ \\
\multicolumn{1}{c}{(1)} & \multicolumn{1}{c}{(2)} & \multicolumn{1}{c}{(3)} & 
\multicolumn{1}{c}{(4)} & \multicolumn{1}{c}{(5)} & \multicolumn{1}{c}{(6)} & 
\multicolumn{1}{c}{(7)} & \multicolumn{1}{c}{(8)} \\
\hline
AO0235$+$164 ($z_{\rm DLA}=0.524$) \dotfill 
                       & Tek\#5/CCD & 0.259 & $U$   & 2700 & 1.22 & 25.2 & 22.9 \\
                       & Tek\#5/CCD & 0.259 & $B$   & 1200 & 1.15 & 25.5 & 22.6 \\
                       & Tek\#5/CCD & 0.259 & $V$   & 2400 & 0.97 & 25.5 & 22.0 \\
                       & Tek\#5/CCD & 0.259 & $R$   & 2100 & 0.85 & 25.7 & 21.2 \\
                       & Tek\#5/CCD & 0.259 & $I$   & 2100 & 0.78 & 24.7 & 20.7 \\
                       &  HST/WFPC2 & 0.100 & F450W & 1200 & 0.20 & 24.7 & 22.5 \\
                       &  HST/WFPC2 & 0.100 & F702W &  600 & 0.20 & 24.4 & 21.0 \\

EX0302$-$2223 ($z_{\rm DLA}=1.0095$) \dotfill 
                       &  HST/WFPC2 & 0.046 & F450W & 2000 & 0.09 & 24.4 & 24.1 \\
                       &  HST/WFPC2 & 0.046 & F702W & 3600 & 0.09 & 25.5 & 23.2 \\
                       & Tek\#5/CCD & 0.259 & $I$   & 8100 & 0.78 & 25.4 & 22.4 \\
                       &   CIRSI    & 0.196 & $J$   & 5700 & 0.52 & 23.8 & 21.7 \\
                       &   CIRSI    & 0.196 & $H$   & 1620 & 0.56 & 23.3 & 21.5 \\

PKS0439$-$433 ($z_{\rm DLA}=0.101$) \dotfill 
                       & Tek\#5/CCD & 0.259 & $U$   & 1800 & 0.85 & 25.3 & 18.8 \\
                       & Tek\#5/CCD & 0.259 & $B$   & 1200 & 0.91 & 25.7 & 18.1 \\
                       & Tek\#5/CCD & 0.259 & $V$   &  900 & 0.78 & 25.2 & 17.5 \\
                       & Tek\#5/CCD & 0.259 & $I$   &  900 & 0.65 & 24.3 & 17.0 \\
                       &   WIRC     & 0.199 & $J$   & 2460 & 0.60 & 23.7 & 16.8 \\
                       &   WIRC     & 0.199 & $K$   & 3240 & 0.46 & 22.5 & 16.9 \\


HE1122$-$1649 ($z_{\rm DLA}=0.6819$) \dotfill 
                       & Tek\#5/CCD & 0.259 & $U$   & 4200 & 0.89 & 26.0 & 23.4 \\
                       & Tek\#5/CCD & 0.259 & $V$   & 2100 & 0.79 & 25.9 & 22.8 \\
                       & Tek\#5/CCD & 0.259 & $I$   & 9600 & 0.70 & 25.7 & 21.4 \\
                       &   CIRSI    & 0.196 & $J$   & 5700 & 0.57 & 24.2 & 20.9 \\
                       &   CIRSI    & 0.196 & $H$   & 5580 & 0.70 & 23.8 & 20.6 \\

PKS1127$-$145 ($z_{\rm DLA}=0.313$) \dotfill 
                       & Tek\#5/CCD & 0.259 & $U$   & 3480 & 0.98 & 26.0 & 21.6 \\
                       & Tek\#5/CCD & 0.259 & $B$   & 3240 & 0.83 & 26.5 & 21.2 \\
                       & Tek\#5/CCD & 0.259 & $V$   & 3600 & 0.67 & 26.4 & 20.3 \\
                       & Tek\#5/CCD & 0.259 & $I$   & 6000 & 0.60 & 25.0 & 19.6 \\
                       &   CIRSI    & 0.196 & $J$   & 4200 & 0.56 & 23.7 & 19.2 \\
                       &   CIRSI    & 0.196 & $H$   & 3300 & 0.56 & 23.9 & 19.2 \\
\hline
\multicolumn{8}{l}{$^a$For a typical Scd galaxy of $M_{AB}(B)=-19.6$, assuming
no luminosity evolution with redshfit.}
\end{tabular}
\end{center}
\end{tiny}


\begin{scriptsize}
\begin{center}
\begin{tabular}{rrrrccccccc}
\multicolumn{11}{c}{Table 2} \\
\multicolumn{11}{c}{Galaxies identified within $15''$ angular radius from 
AO0235$+$164 ($z_{\rm em}=0.940$; $z_{\rm DLA}=0.524$)} \\
\hline
\hline
\multicolumn{1}{c}{ID} & \multicolumn{1}{c}{$\Delta\alpha$ ($''$)}  & 
\multicolumn{1}{c}{$\Delta\delta$ ($''$)} & \multicolumn{1}{c}{$\Delta\theta$ 
($''$)} & $z_{\rm phot}$  & $z_{\rm spec}$ & 
\multicolumn{1}{c}{$U_{AB}$} & \multicolumn{1}{c}{$B_{AB}$} &
\multicolumn{1}{c}{$V_{AB}$} & \multicolumn{1}{c}{$R_{AB}$} & 
\multicolumn{1}{c}{$I_{AB}$} \\
\multicolumn{1}{c}{(1)} & \multicolumn{1}{c}{(2)} & \multicolumn{1}{c}{(3)} & 
\multicolumn{1}{c}{(4)} & (5) & (6) & \multicolumn{1}{c}{(7)} & 
\multicolumn{1}{c}{(8)} & \multicolumn{1}{c}{(9)} & \multicolumn{1}{c}{(10)} & 
\multicolumn{1}{c}{(11)} \\
\hline
  1 &    1.1 & 0.0 &  1.1 &  ... & 0.524$^{a}$ & ... & ... & ... & $\ge 21.36$$^e$ & ... \\ 
  2 &    0.2 & $-$2.1 &  2.1 &  0.52 & 0.524$^{a,b,c}$ & 21.06$\pm$0.03 & 20.84$\pm$0.01 & 20.61$\pm$0.01 & 20.42$\pm$0.01 & 20.22$\pm$0.02 \\ 
  3 & $-$5.9 & $-$2.6 &  6.5 &  0.32 & 0.525$^{a,d}$ & 24.42$\pm$0.54 & 24.16$\pm$0.30 & 22.48$\pm$0.06 & 21.54$\pm$0.03 & 20.85$\pm$0.03 \\ 
  4 & $-$6.9 & $-$7.0 &  9.8 &  0.68 &    ...    & 23.97$\pm$0.38 & 24.70$\pm$0.53 & 23.40$\pm$0.15 & 22.54$\pm$0.06 & 21.21$\pm$0.04 \\ 
\hline
\multicolumn{11}{l}{$^a$Yanny \etal\ 1989; $^b$ Cohen \etal\ 1987;
$^c$Guillemin \& Bergeron 1997; $^d$Stickel \etal\ 1988; } \\
\multicolumn{11}{l}{$^e$Magnitude limit was estimated based on a WFPC2/F702W 
image.} \\
\end{tabular}
\end{center}
\end{scriptsize}


\begin{scriptsize}
\begin{center}
\begin{tabular}{rrrrccccccc}
\multicolumn{11}{c}{Table 3} \\
\multicolumn{11}{c}{Galaxies identified within $15''$ angular radius from 
EX0302$-$2223 ($z_{\rm em}=1.400$; $z_{\rm DLA}=1.0095$)} \\
\hline
\hline
\multicolumn{1}{c}{ID} & \multicolumn{1}{c}{$\Delta\alpha$ ($''$)}  & 
\multicolumn{1}{c}{$\Delta\delta$ ($''$)} & \multicolumn{1}{c}{$\Delta\theta$ 
($''$)} & $z_{\rm phot}$  & $z_{\rm spec}$ & 
\multicolumn{1}{c}{F450W} & \multicolumn{1}{c}{F702W} &
\multicolumn{1}{c}{$I_{AB}$} & \multicolumn{1}{c}{$J_{AB}$} & 
\multicolumn{1}{c}{$H_{AB}$} \\
\multicolumn{1}{c}{(1)} & \multicolumn{1}{c}{(2)} & \multicolumn{1}{c}{(3)} & 
\multicolumn{1}{c}{(4)} & (5) & (6) & \multicolumn{1}{c}{(7)} & 
\multicolumn{1}{c}{(8)} & \multicolumn{1}{c}{(9)} & \multicolumn{1}{c}{(10)} & 
\multicolumn{1}{c}{(11)} \\
\hline
    1 & $-$1.1 & $-$0.3 &  1.1 & ... & ... & 23.68$\pm$0.17& 23.88$\pm$0.08& ... & ... & ... \\ 
    2 & $-$2.4 &    1.2 &  2.7 & ... & ... & 24.40$\pm$0.37& 23.78$\pm$0.07& ... & ... & ... \\ 
    3 &    0.2 &    3.4 &  3.4 &  0.96 & ... & 24.66$\pm$0.42& 24.04$\pm$0.09& 23.13$\pm$0.04& 23.69$\pm$0.30 & 24.50$\pm$0.87 \\ 
    4 &    1.0 &    3.6 &  3.7 &  ...  & ... & 24.88$\pm$0.57& 23.96$\pm$0.09& 23.81$\pm$0.09& 23.13$\pm$0.20 & 22.90$\pm$0.24 \\ 
3$+$4 &    0.6 &    3.5 &  3.6 &  0.96 & ... & 24.01$\pm$0.34& 23.25$\pm$0.06& 22.67$\pm$0.04& 22.62$\pm$0.17 & 22.68$\pm$0.25 \\ 
    5 & $-$2.7 & $-$7.1 &  7.6 &  1.05 & 1.000$^a$ & 24.74$\pm$0.54& 22.99$\pm$0.04& 21.67$\pm$0.01& 20.71$\pm$0.02 & 20.79$\pm$0.04 \\ 
    6 & $-$8.3 &   11.8 & 14.4 &  0.70 & 0.663$^a$ & 23.88$\pm$0.38& 21.59$\pm$0.02& 21.10$\pm$0.01& 20.57$\pm$0.03 & 20.46$\pm$0.04 \\ 
    7 &   14.8 & $-$0.7 & 14.8 &  0.04 & 0.118$^b$ & 19.62$\pm$0.02& 18.73$\pm$0.01& 18.43$\pm$0.01& 18.06$\pm$0.01 & 17.88$\pm$0.01 \\ 
\hline
\multicolumn{11}{l}{$^a$Guillemin \& Bergeron 1997; $^b$Le Brun \etal\ 1997.} \\
\end{tabular}
\end{center}
\end{scriptsize}

\begin{tiny}
\begin{center}
\begin{tabular}{rrrrcccccccc}
\multicolumn{12}{c}{Table 4} \\
\multicolumn{12}{c}{Galaxies identified within $15''$ angular radius from 
PKS0439$-$433 ($z_{\rm em}=0.593$; $z_{\rm DLA}=0.101$)} \\
\hline
\hline
\multicolumn{1}{c}{ID} & \multicolumn{1}{c}{$\Delta\alpha$ ($''$)}  & 
\multicolumn{1}{c}{$\Delta\delta$ ($''$)} & \multicolumn{1}{c}{$\Delta\theta$ 
($''$)} & $z_{\rm phot}$  & $z_{\rm spec}$ & 
\multicolumn{1}{c}{$U_{AB}$} & \multicolumn{1}{c}{$B_{AB}$} &
\multicolumn{1}{c}{$V_{AB}$} & \multicolumn{1}{c}{$I_{AB}$} & 
\multicolumn{1}{c}{$J_{AB}$} & \multicolumn{1}{c}{$K_{AB}$} \\
\multicolumn{1}{c}{(1)} & \multicolumn{1}{c}{(2)} & \multicolumn{1}{c}{(3)} & 
\multicolumn{1}{c}{(4)} & (5) & (6) & \multicolumn{1}{c}{(7)} & 
\multicolumn{1}{c}{(8)} & \multicolumn{1}{c}{(9)} & \multicolumn{1}{c}{(10)} & 
\multicolumn{1}{c}{(11)} & \multicolumn{1}{c}{(12)} \\
\hline
  1 & $-$0.5 &     3.9 &  3.9 &  0.09 & 0.101$^{a,b}$ & 19.31$\pm$0.01 & 18.37
$\pm$0.01 & 17.72$\pm$0.01 & 17.12$\pm$0.01 & 16.55$\pm$0.01 & 16.66$\pm$0.01 \\ 
  2 & $-$4.2 & $-$10.1 & 10.9 &  0.00 & ... & 21.74$\pm$0.04 & 19.67$\pm$0.01 &
18.76$\pm$0.01 & 17.88$\pm$0.01 & 17.47$\pm$0.01 & 17.85$\pm$0.02 \\ 
  3 &    5.6 &    12.0 & 13.3 &  0.83 & ... & 25.33$\pm$0.47 & 25.78$\pm$0.38 &
25.02$\pm$0.32 & 22.21$\pm$0.08 & 21.76$\pm$0.09 & 20.36$\pm$0.07 \\ 
\hline
\multicolumn{11}{l}{$^a$Petitjean \etal\ (1996); $^b$Chen, Kennicutt, \& Rauch 
(2003, in preparation).} \\
\end{tabular}
\end{center}
\end{tiny}

\begin{scriptsize}
\begin{center}
\begin{tabular}{rrrrccccccc}
\multicolumn{11}{c}{Table 5} \\
\multicolumn{11}{c}{Galaxies identified within $15''$ angular radius from 
HE1122$-$168 ($z_{\rm em}=2.400$; $z_{\rm DLA}=0.6819$)} \\
\hline
\hline
\multicolumn{1}{c}{ID} & \multicolumn{1}{c}{$\Delta\alpha$ ($''$)}  & 
\multicolumn{1}{c}{$\Delta\delta$ ($''$)} & \multicolumn{1}{c}{$\Delta\theta$ 
($''$)} & $z_{\rm phot}$  & $z_{\rm spec}$ & 
\multicolumn{1}{c}{$U_{AB}$} & \multicolumn{1}{c}{$V_{AB}$} &
\multicolumn{1}{c}{$I_{AB}$} & \multicolumn{1}{c}{$J_{AB}$} & 
\multicolumn{1}{c}{$H_{AB}$} \\
\multicolumn{1}{c}{(1)} & \multicolumn{1}{c}{(2)} & \multicolumn{1}{c}{(3)} & 
\multicolumn{1}{c}{(4)} & (5) & (6) & \multicolumn{1}{c}{(7)} & 
\multicolumn{1}{c}{(8)} & \multicolumn{1}{c}{(9)} & \multicolumn{1}{c}{(10)} & 
\multicolumn{1}{c}{(11)} \\
\hline
  1 &  0.8 & $-$3.5 &  3.6 &  0.69 & ... & 23.19$\pm$0.08 & 23.01$\pm$0.06 & 22.33$\pm$0.03 & 22.10$\pm$0.10 & 21.72$\pm$0.09 \\ 
  2 & $-$2.8 & $-$6.5 &  7.1 &  1.24 & ... & 24.41$\pm$0.20 & 24.28$\pm$0.15 & 23.56$\pm$0.08 & $>24.95$ & 22.87$\pm$0.21 \\ 
  3 &  0.2 &  7.4 &  7.4 &  1.05 & ... & 25.35$\pm$0.48 & 25.20$\pm$0.35 & 23.36$\pm$0.07 & 22.36$\pm$0.11 & 22.39$\pm$0.15 \\ 
  4 &  3.7 & $-$6.6 &  7.5 &  0.72 & ... & 26.22$\pm$0.68& 25.72$\pm$0.38 & 24.33$\pm$0.11 & 26.06$\pm$1.86 & 23.49$\pm$0.25 \\ 
  5 & $-$1.2 & $-$8.5 &  8.6 &  0.00 & ... & 24.42$\pm$0.20 & 23.87$\pm$0.10 & 23.07$\pm$0.05 & 23.16$\pm$0.21 & 23.72$\pm$0.51 \\ 
  6 &  3.1 & 10.0 & 10.4 &  0.63 & ... & 24.82$\pm$0.27 & 24.35$\pm$0.15 & 23.34$\pm$0.06 & 23.47$\pm$0.27 & 22.63$\pm$0.18 \\ 
  7 & 10.7 & $-$1.9 & 10.9 &  0.26 & ... & 28.11$\pm$5.36 & 24.98$\pm$0.23 & 23.12$\pm$0.05 & 22.51$\pm$0.11 & 22.72$\pm$0.17 \\ 
  8 & $-$6.1 &  9.6 & 11.4 &  0.43 & ... & $>25.83$ & 23.25$\pm$0.09 & 21.10$\pm$0.01 & 20.47$\pm$0.03 & 19.97$\pm$0.02 \\ 
  9 & 10.6 &  3.2 & 11.0 &  0.35 & ... & $>26.06$ & 23.87$\pm$0.12 & 21.97$\pm$0.02 & 21.28$\pm$0.04 & 21.01$\pm$0.05 \\ 
  10 & $-$3.0 & $-$12.1 & 12.4 &  1.22 & ... & 25.14$\pm$0.33 & 25.05$\pm$0.25 & 23.06$\pm$0.04 & 21.76$\pm$0.05 & 21.47$\pm$0.05 \\ 
\hline
\end{tabular}
\end{center}
\end{scriptsize}

\begin{tiny}
\begin{center}
\begin{tabular}{rrrrcccccccc}
\multicolumn{12}{c}{Table 6} \\
\multicolumn{12}{c}{Galaxies identified within $15''$ angular radius from 
PKS1127$-$145 ($z_{\rm em}=1.187$; $z_{\rm DLA}=0.313$)} \\
\hline
\hline
\multicolumn{1}{c}{ID} & \multicolumn{1}{c}{$\Delta\alpha$ ($''$)}  & 
\multicolumn{1}{c}{$\Delta\delta$ ($''$)} & \multicolumn{1}{c}{$\Delta\theta$ 
($''$)} & $z_{\rm phot}$  & $z_{\rm spec}$ & 
\multicolumn{1}{c}{$U_{AB}$} & \multicolumn{1}{c}{$B_{AB}$} &
\multicolumn{1}{c}{$V_{AB}$} & \multicolumn{1}{c}{$I_{AB}$} & 
\multicolumn{1}{c}{$J_{AB}$} & \multicolumn{1}{c}{$H_{AB}$} \\
\multicolumn{1}{c}{(1)} & \multicolumn{1}{c}{(2)} & \multicolumn{1}{c}{(3)} & 
\multicolumn{1}{c}{(4)} & (5) & (6) & \multicolumn{1}{c}{(7)} & 
\multicolumn{1}{c}{(8)} & \multicolumn{1}{c}{(9)} & \multicolumn{1}{c}{(10)} & 
\multicolumn{1}{c}{(11)} & \multicolumn{1}{c}{(12)} \\
\hline
 1 & $-$1.9 &  3.3 &  3.8 &  0.27 & ... & 24.00$\pm$0.15 & 23.90$\pm$0.07 & 23.16$\pm$0.04 & 22.36$\pm$0.05 & 22.22$\pm$0.15 & 22.27$\pm$0.13 \\ 
 2 & $-$3.8 &  0.4 &  3.8 &  0.33 & 0.312$^a$ & 23.32$\pm$0.07 & 23.32$\pm$0.03 & 22.68$\pm$0.02 & 22.14$\pm$0.04 & 21.97$\pm$0.10 & 22.60$\pm$0.15 \\ 
 3 &  5.4 & $-$1.4 &  5.6 &  0.12 & ... & 23.06$\pm$0.07 & 22.94$\pm$0.03 & 22.26$\pm$0.02 & 21.70$\pm$0.04 & 21.95$\pm$0.14 & 21.70$\pm$0.09 \\ 
 4 & $-$3.0 &  7.1 &  7.7 &  0.75 & ... & 25.31$\pm$0.32 & 24.57$\pm$0.09 & 24.56$\pm$0.10 & 23.36$\pm$0.08 & 23.76$\pm$0.40 & 22.99$\pm$0.17 \\ 
 5 &  8.0 &  1.9 &  8.2 &  0.53 & ... & 23.61$\pm$0.10 & 23.48$\pm$0.04 & 23.02$\pm$0.03 & 22.29$\pm$0.05 & 21.58$\pm$0.08 & 21.31$\pm$0.05 \\ 
 6 &  9.0 &  3.9 &  9.8 &  0.26 & 0.313$^b$ & 21.47$\pm$0.04 & 21.03$\pm$0.01 & 19.90$\pm$0.01 & 19.09$\pm$0.01 & 18.59$\pm$0.01 & 18.30$\pm$0.01 \\ 
 7 &  2.7 & 11.9 & 12.2 &  0.00 & ... & 25.30$\pm$0.36 & 23.46$\pm$0.04 & 22.57$\pm$0.02 & 21.67$\pm$0.02 & 21.41$\pm$0.07 & 21.48$\pm$0.06 \\ 
 8 & $-$3.1 & 14.6 & 15.0 &  0.95 & ... & 25.37$\pm$0.29 & 25.26$\pm$0.12 & 24.94$\pm$0.11 & 23.89$\pm$0.10 & 24.54$\pm$0.70 & 24.66$\pm$0.65 \\ 
\hline
\multicolumn{12}{l}{$^a$Lane \etal\ (1998); $^b$Bergeron \& Boiss\'e (1991).} \\
\end{tabular}
\end{center}
\end{tiny}

\begin{scriptsize}
\begin{center}
\begin{tabular}{p{1.5 in}crcrrrcc}
\multicolumn{9}{c}{Table 7} \\
\multicolumn{9}{c}{Properties of known DLA galaxies at $z\le 1$} \\
\hline
\hline
 & & & $\log N(\hI)$ & \multicolumn{1}{c}{$\Delta\theta$} & 
\multicolumn{1}{c}{$\rho$} & & $M_{AB}(B)$ & \\
\multicolumn{1}{c}{QSO} & $z_{\rm em}$ & $z_{\rm abs}$ & (cm$^{-2}$) & 
\multicolumn{1}{c}{(arcsec)} & \multicolumn{1}{c}{($h^{-1}$ kpc)} & 
\multicolumn{1}{c}{$AB$} & $-5\,\log\,h$${^a}$ & Morphology \\
\multicolumn{1}{c}{(1)} & (2) & (3) & (4) & \multicolumn{1}{c}{(5)} & 
\multicolumn{1}{c}{(6)} & \multicolumn{1}{c}{(7)} & (8) & (9) \\
\hline
TON 1480 \dotfill & 0.614 & 0.0036 & 20.34 & 114.0 & 5.94 & $B=11.5$ & $-18.7$ & S0${^b}$ \\
HS1543$+$5921 \dotfill & 0.807 & 0.009 & 20.35 & 2.4 & 0.31 & $R=16.5$ & $-15.3$ & LSB${^c}$ \\
\hline
LBQS0058$+$0155 \dotfill & 1.954 & 0.613 & 20.08 & 1.2 & 5.67 & $R=23.7$ & $-17.6$ & disk \\
AO0235$+$164 \dotfill & 0.940 & 0.524 & 21.70 & 2.1 &  9.37 & $I=20.2$ & $-20.3$ & compact \\
                      &       &       &       & 6.4 & 28.03 & $20.9$ & $-19.7$ & compact \\
EX0302$-$2223 \dotfill & 1.400 & 1.0095 & 20.36 & 3.3 & 18.65 & $I=22.7$ & $-19.3$ & Irr \\ 
PKS0439$-$433 \dotfill & 0.593 & 0.101 & 20.00 & 3.9 & 5.13 & $I=17.2$ & $-19.6$ & disk \\
Q0738$+$313 \dotfill & 0.635 & 0.2212 & 20.90 & 5.7 & 14.23 & $I=20.9$ & $-17.7$ & compact \\
B2\,0827$+$243 \dotfill & 0.939 & 0.525 & 20.30 & 5.8 & 25.42 & $R=21.0$ & $-20.0$ & disk \\
HE1122$-$1649 \dotfill & 2.400 & 0.681 & 20.45 & 3.6 & 17.66 & $I=22.4$ & $-18.8$ & compact \\ 
PKS1127$-$145\dotfill & 1.187 & 0.313 & 21.71 &  3.8 & 12.13 & $R=22.4$ & $-16.8$ & Irr \\
                       &       &       &       &  3.8 & 12.13 &     22.1 & $-17.4$ & compact \\
                       &       &       &       &  9.8 & 31.59 &     19.1 & $-20.1$ & disk \\ 
                       &       &       &       & 17.5 & 56.18 &     18.9 & $-20.3$ & disk \\
PKS1629$+$120 \dotfill & 1.795 & 0.531 & 20.70 & 3.0 & 13.24 & $R=21.6$ & $-19.2$ & disk \\
\hline
Q0738$+$313 \dotfill & 0.635 & 0.0912 & 21.18 & ... & ... & $K>17.8$ & $>-18.8$ & \\
PKS1622$+$23\dotfill & 0.927 & 0.6563 & 20.36 & ... & ... & $R>24.5$ & $>-16.9$ & \\
\hline
\multicolumn{9}{l}{$^a$$M_{{AB}_*}(B)-5\log h = -19.6$; $^b$Nilson (1973); 
$^c$Bowen \etal\ (2001).}
\end{tabular}
\end{center}
\end{scriptsize}


\newpage

\begin{figure}
\plotone{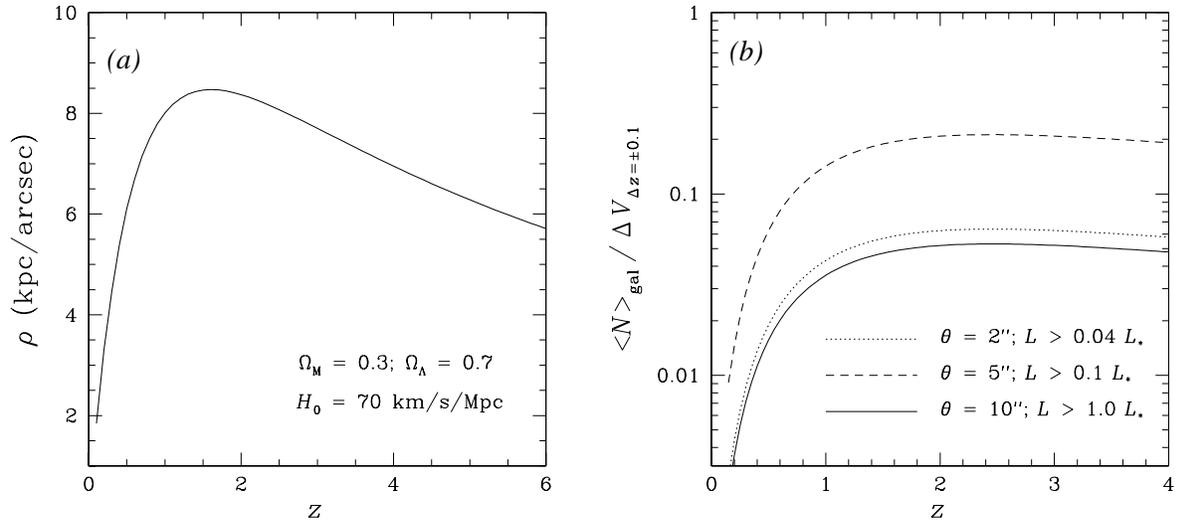}
\caption[]{(a) The corresponding projected distance in kpc per one arcsec 
angular separation versus redshift; (b) The number of random galaxies expected
in a cylinder of $\theta$ in angular radius and $\Delta z$ in length.  The 
curves are calculated by integrating over a galaxy luminosity function--taken 
from the Autofib survey by Ellis \etal\ (1996)--to different luminosity limits 
and different angular radii.}
\end{figure}

\newpage

\begin{figure}
\epsscale{0.75}
\plotone{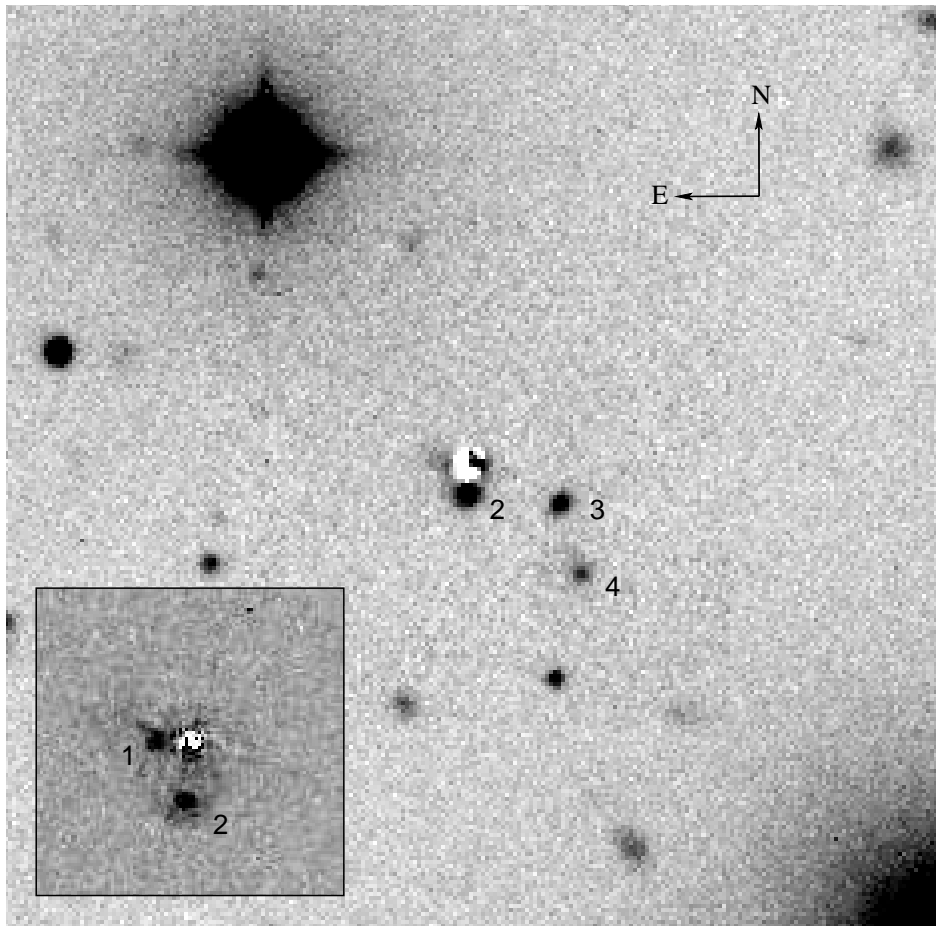}
\caption[]{The field around AO0235$+$164 obtained with the Tek\#5 CCD camera 
using the $I$ filter.  The image is $60''$ on a side.  The light from the
background QSO has been subtracted using an empirical PSF determined from 26
stars in the image frame.  The inset shows a close-up image ($10''$ on a side)
of the immediate vicinity of the QSO obtained with HST using the F702W filter.
Object \#1 at ($+1.1''$,$0.0''$) to the QSO is clearly seen in the HST image, 
but is not resolved in the groud-based ones.}
\end{figure}

\newpage

\begin{figure}
\plotone{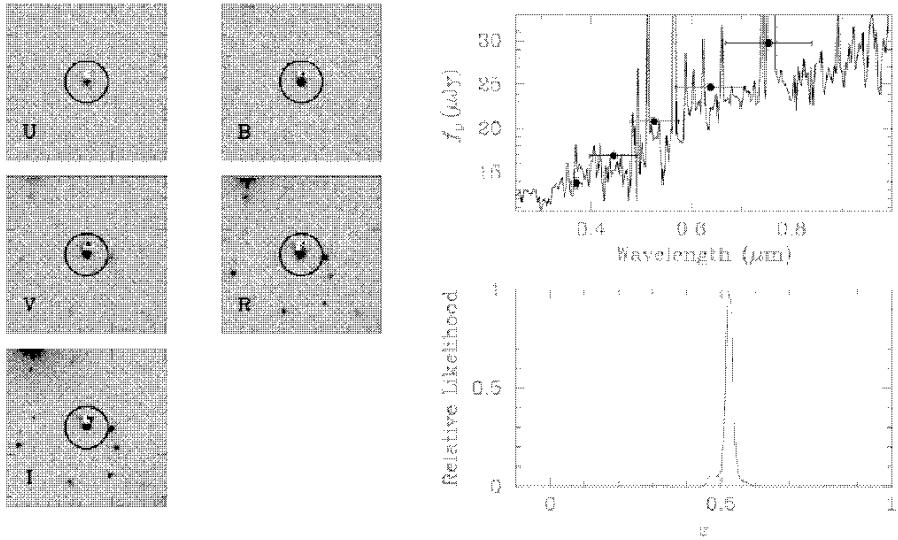}
\caption[]{The DLA galaxy (object 2) at $z\approx 0.524$ in the field of 
AO0235$+$164.  Individual images on the left show this system in different 
bandpasses as indicated in the lower-left corner of each panel.  The dimension
of these image clips is $\approx 40''$ on a side.  The top panel on the right 
shows the observed SED established based on the five photometric measurements 
(solid points) of the galaxy, in comparison to the best-fit starburst template 
solid line).  The bottom panel on the right shows the redshift likelihood 
function with a best-fit photometric redshift at $z_{\rm phot} = 0.52$.}
\end{figure}

\newpage

\begin{figure}
\plotone{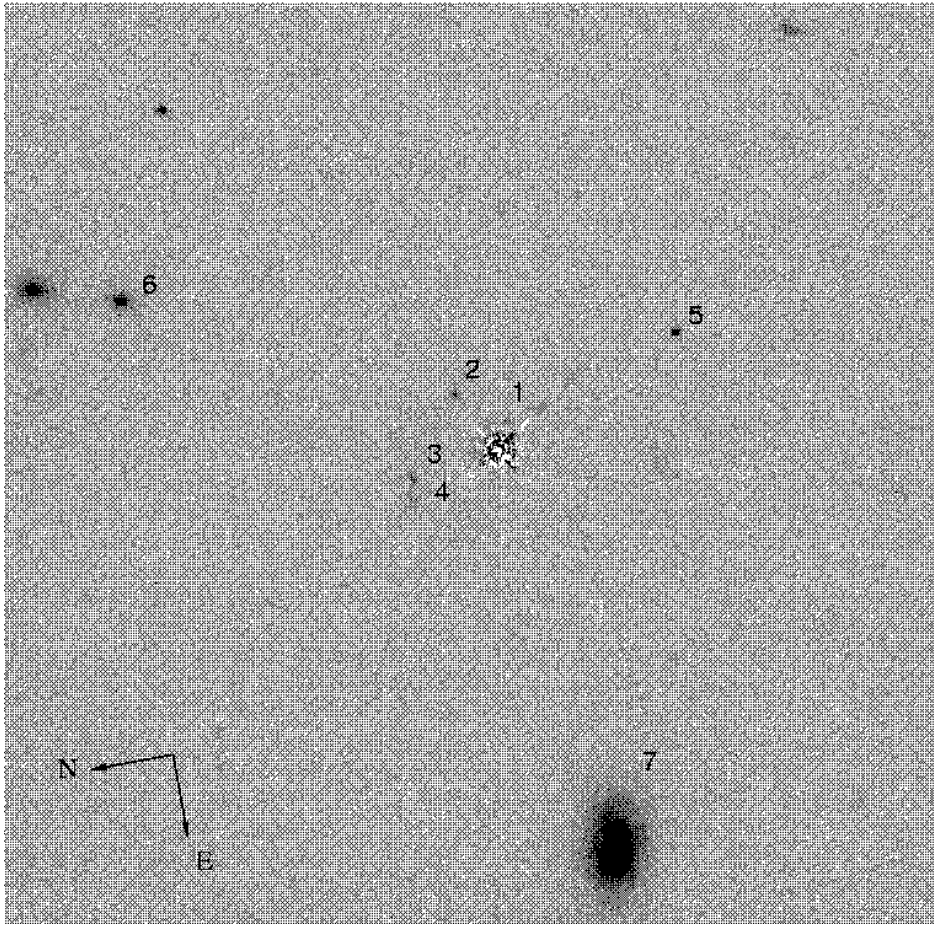}
\caption[]{The field around EX0302$-$2223 obtained with the Planetary Camera 
using the F702W filter.  The image is $33''$ on a side.  The light from the
background QSO has been subtracted using a model PSF determined from the 
Tiny Tim software (Krist \& Hook 1997).}
\end{figure}

\newpage

\begin{figure}
\plotone{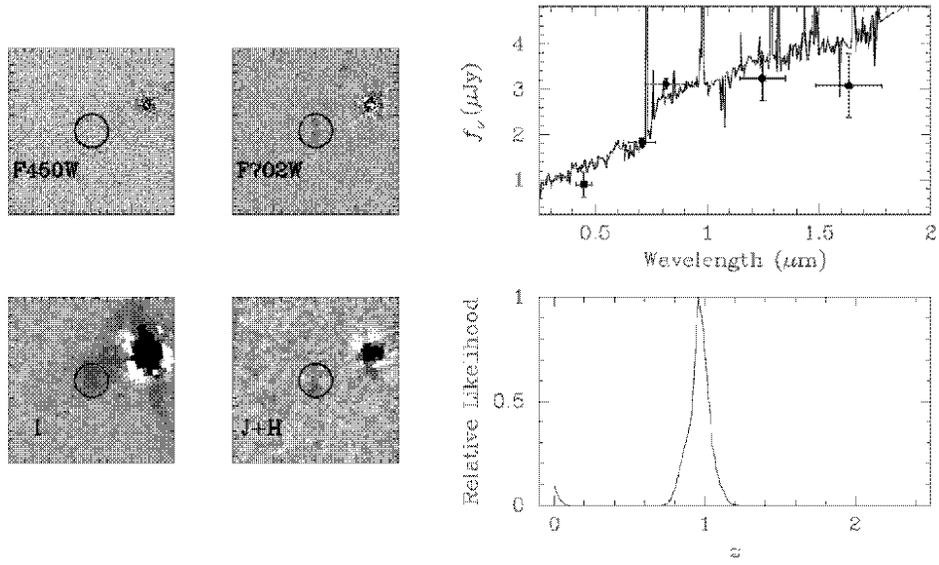}
\caption[]{The DLA galaxy (objects \#3$+$4) at $z\approx 1$ in the field of 
EX0302$-$2223.  The arrangement of the panels is the same as in Figure 3 with 
the dimension of individual images being $9''$ on a side. The best-fit template
is starburst and the best-fit photometric redshift is at $z_{\rm phot}=0.96$.}
\end{figure}

\newpage

\begin{figure}
\epsscale{0.75}
\plotone{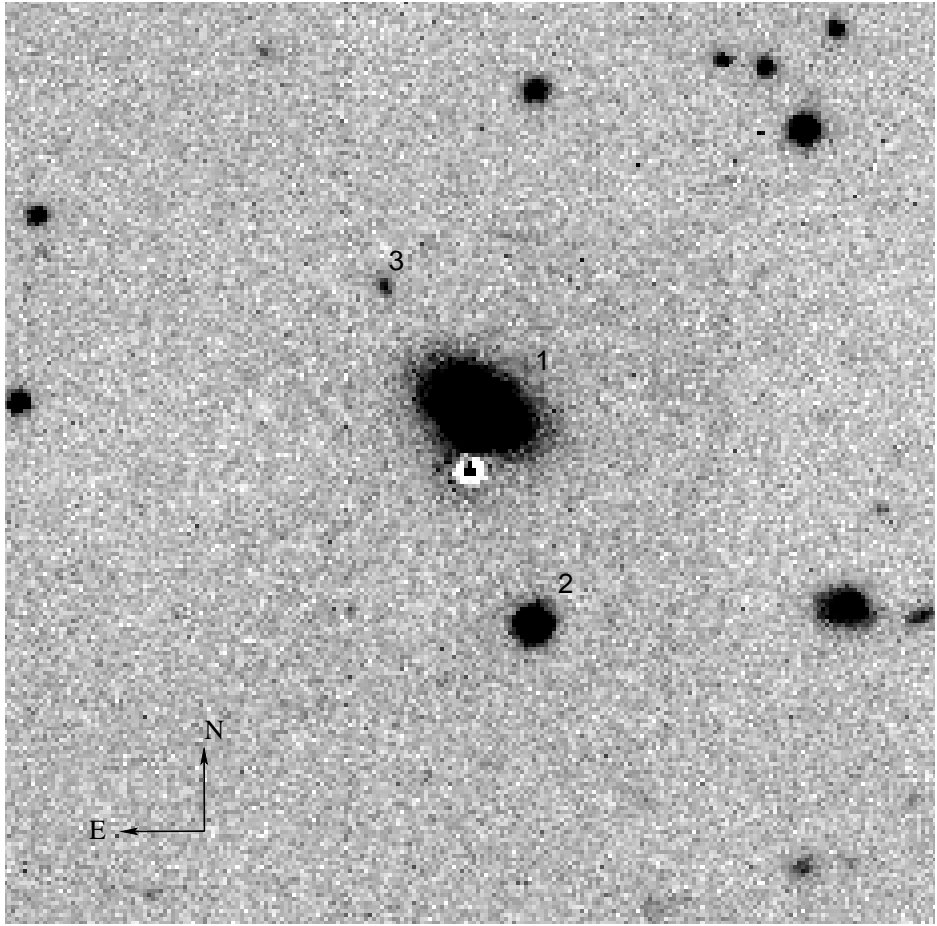}
\caption[]{The field around PKS0439$-$433 obtained with the Tek\#5 CCD camera 
using the $I$ filter.  The image is $60''$ on a side.  The light from the
background QSO has been subtracted using an empirical PSF determined from 10
stars in the image frame.}
\end{figure}

\newpage

\begin{figure}
\plotone{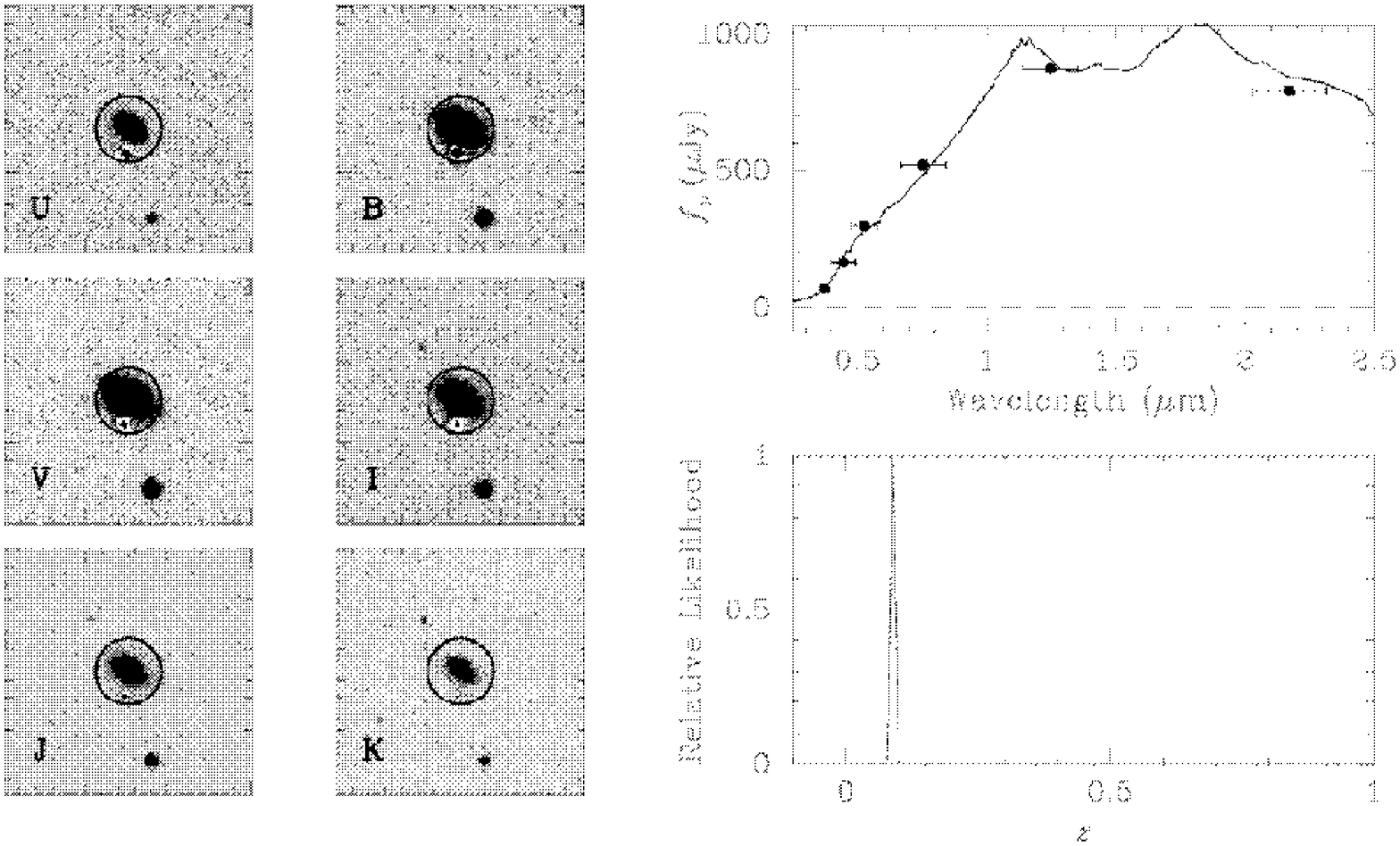}
\caption[]{The DLA galaxy (object 1) at $z\approx 0.1$ in the field of 
PKS0439$-$433.  The arrangement of the panels is the same as in Figure 3 with 
the dimension of individual images being $40''$ on a side. The best-fit 
template is Sab and the best-fit photometric redshift is at $z_{\rm phot} =
0.09$.}
\end{figure}

\newpage

\begin{figure}
\epsscale{0.75}
\plotone{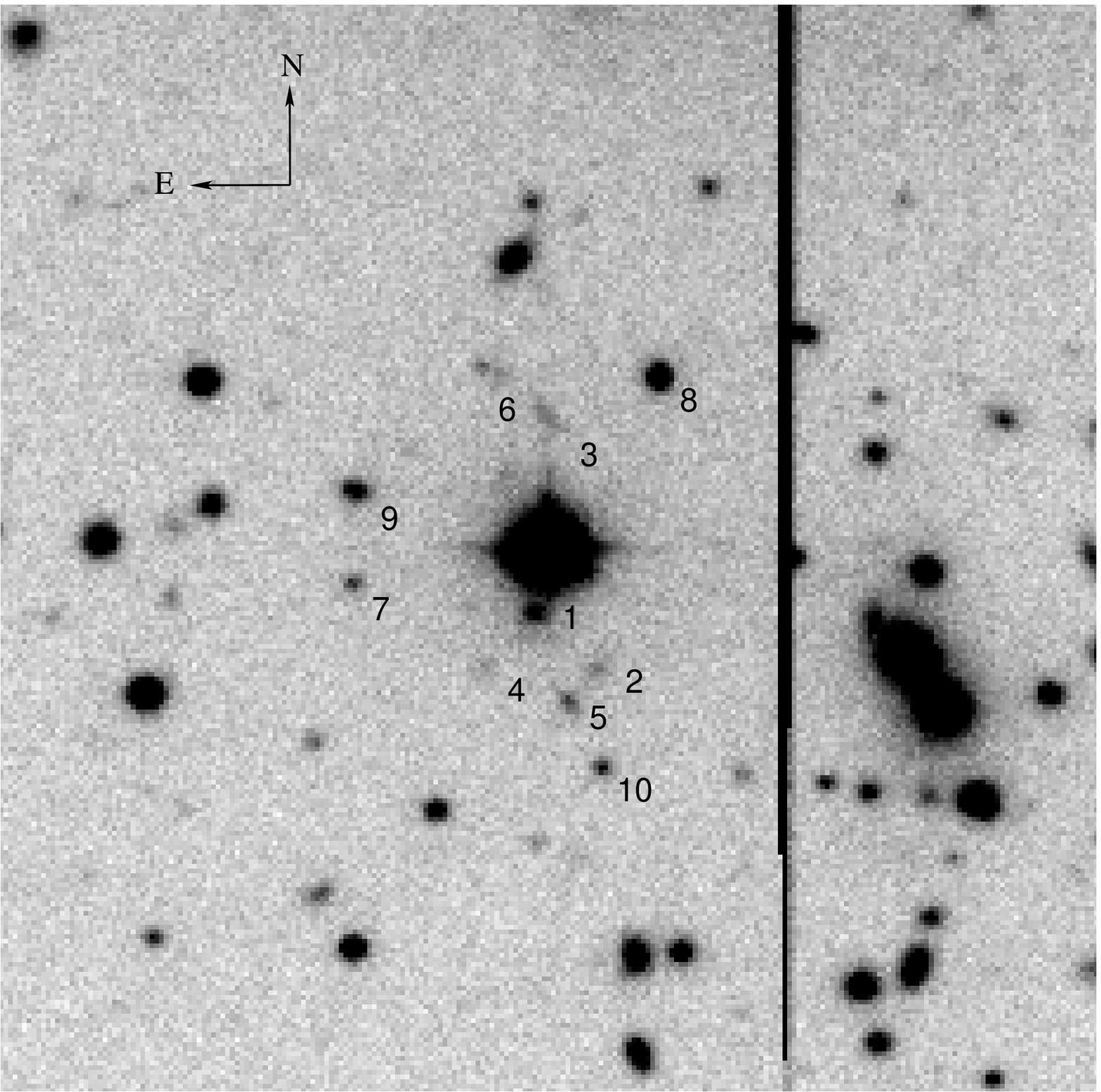}
\caption[]{The field around HE1122$-$1649 obtained with the Tek\#5 CCD camera 
using the $I$ filter.  The QSO is at the center of the image, which is $60''$ 
on a side.}
\end{figure}

\newpage

\begin{figure}
\plotone{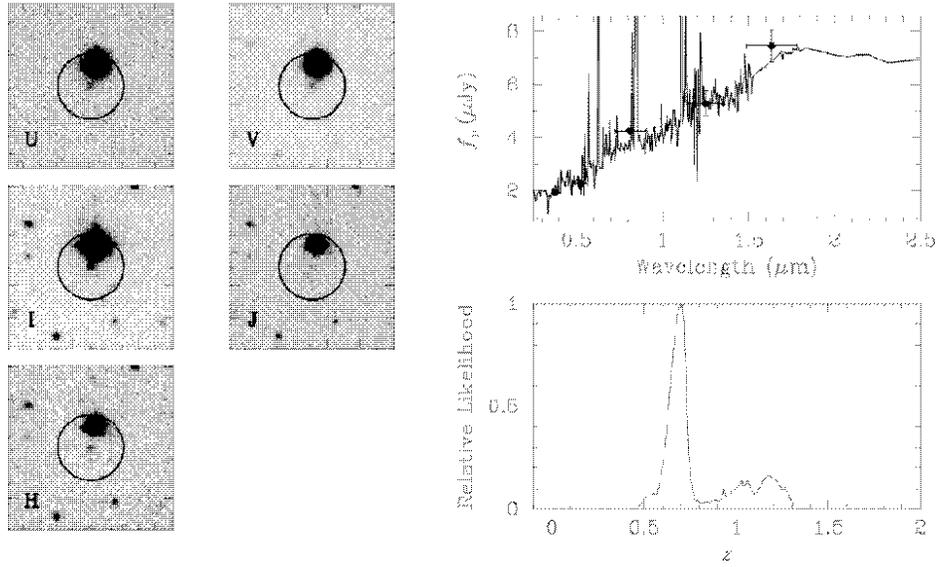}
\caption[]{The DLA galaxy (object 1) at $z\approx 0.68$ in the field of 
HE1122$-$1649.  The arrangement of the panels is the same as in Figure 3 with 
the dimension of individual images being $25''$ on a side. The best-fit 
template is starburst and the best-fit photometric redshift is at $z_{\rm phot}
=0.69$.}
\end{figure}

\newpage

\begin{figure}
\epsscale{0.75}
\plotone{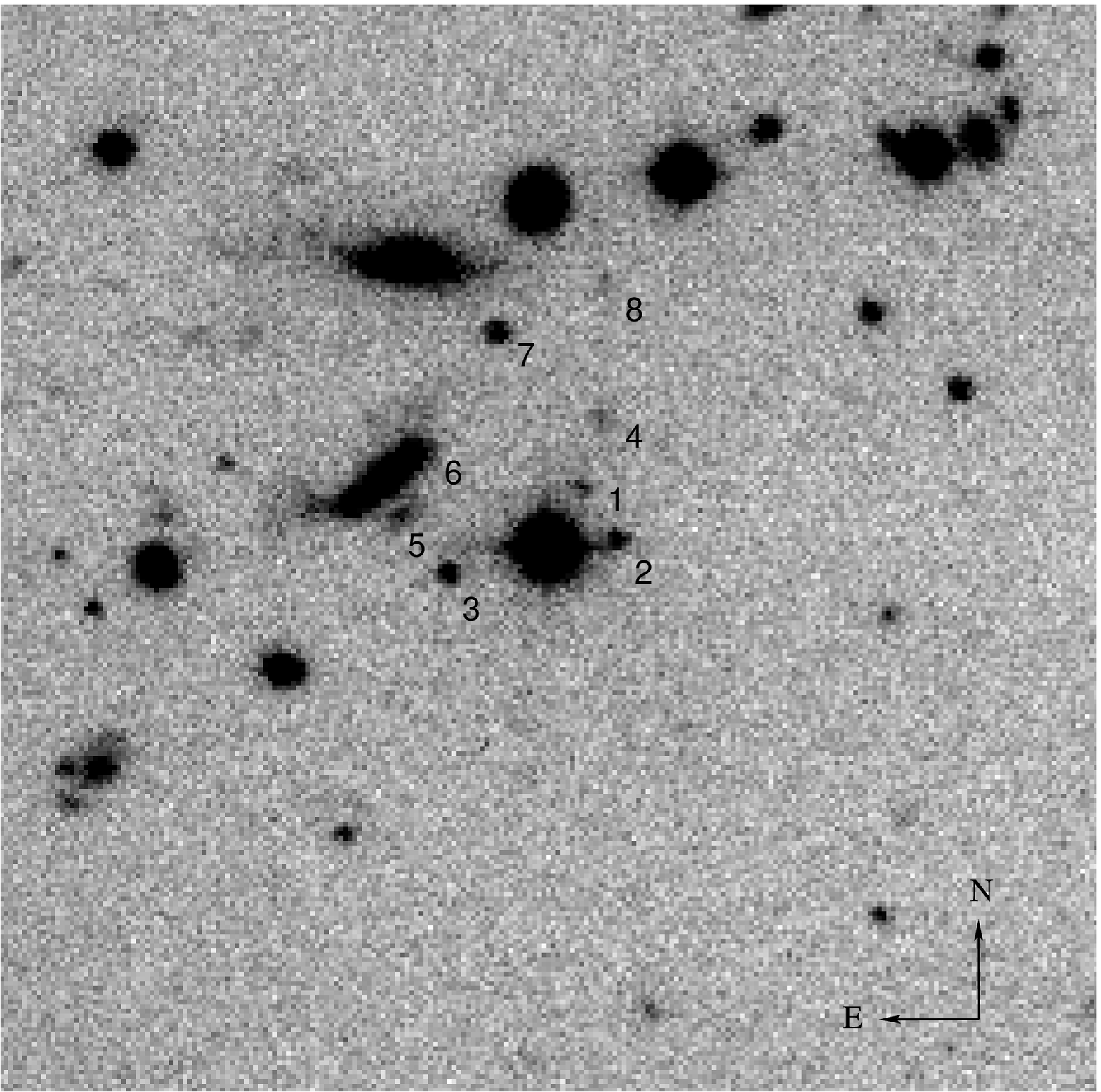}
\caption[]{The field around PKS1127$-$145 obtained with the Tek\#5 CCD camera 
using the $I$ filter.  The QSO is at the center of the image, which is $60''$ 
on a side.}
\end{figure}

\newpage

\begin{figure}
\plotone{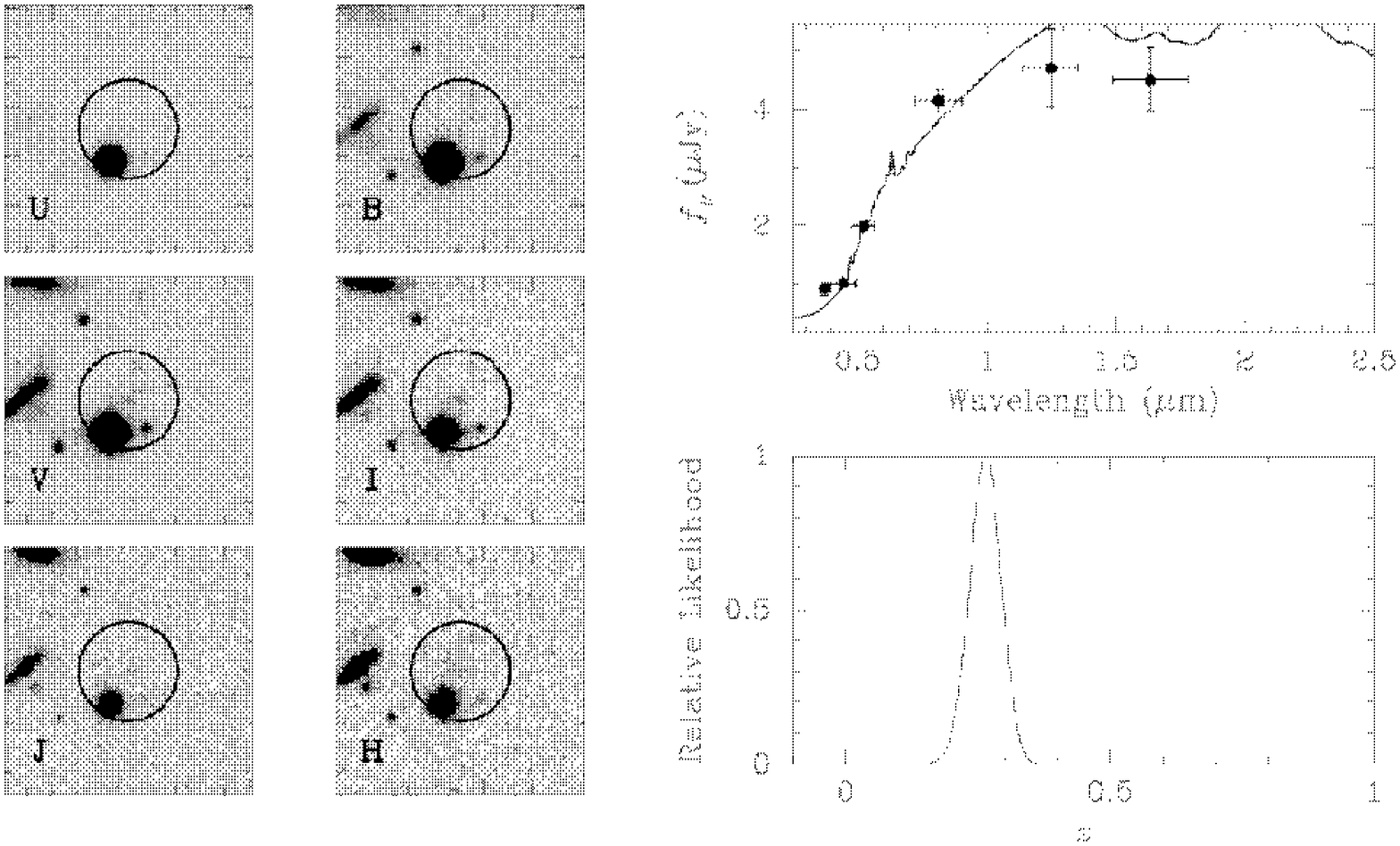}
\caption[]{The DLA galaxy (object 1) at $z\approx 0.313$ in the field of 
PKS1127$-$145.  The arrangement of the panels is the same as in Figure 3 with 
the dimension of individual images being $25''$ on a side. The best-fit 
template is Scd and the best-fit photometric redshift is at $z_{\rm phot} =
0.27$.}
\end{figure}

\newpage

\begin{figure}
\plotone{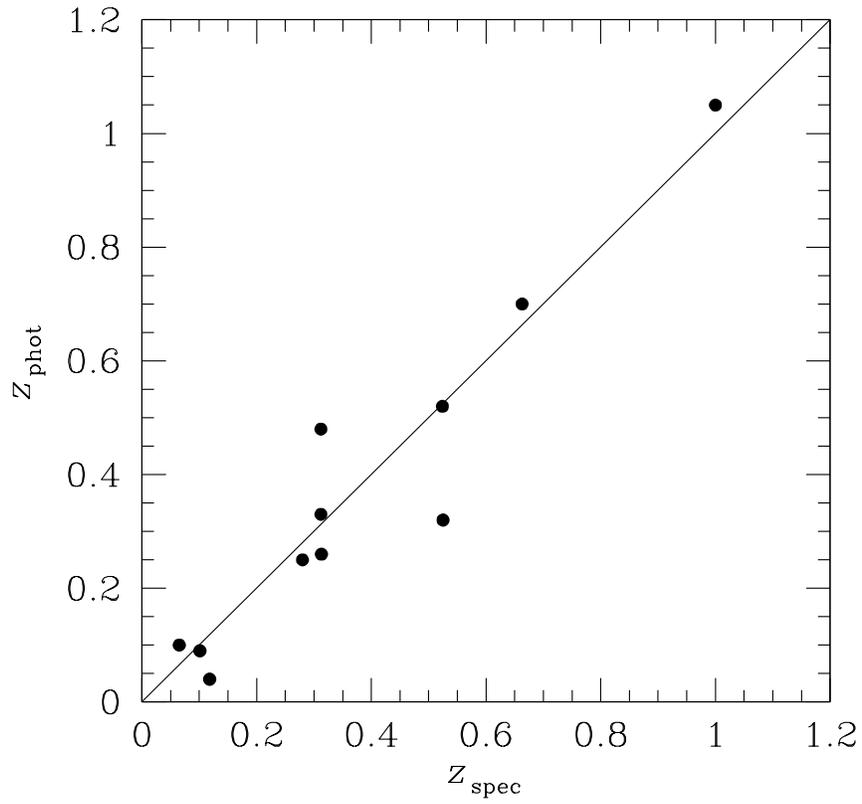}
\caption{Comparison of our photometric redshift measurements and previously
known spectroscopic redshifts for 11 galaxies in the five DLA fields that we 
have surveyed.}
\end{figure}

\newpage

\begin{figure}
\epsscale{1.01}
\plotone{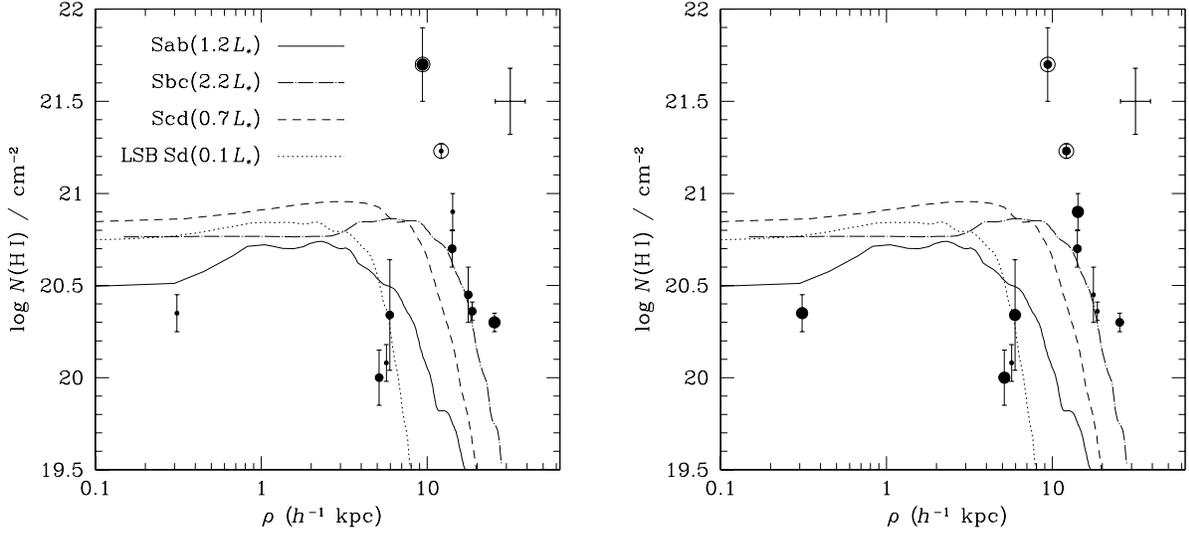}
\caption[]{Neutral hydrogen column density $N(\hI)$ distribution versus galaxy
impact parameter $\rho$ from 11 DLA galaxies (points), in comparison to the 
mean \hI\ profiles of nearby galaxies of different morphological type and
intrinsic luminosity shown in curves (the mean \hI\ profiles for Sab, Sbc, and
Scd galaxies were digitized from Cayatte \etal\ 1994; the curve for LSB Sd was 
provided by Uson \& Matthews 2003).  The scatter of $N(\hI)$ at the Holmberg 
radii and the scatter of the neutral gaseous extent at $N(\hI)=10^{20}$ \cmjj\ 
observed in the 21-cm data for Sd-type galaxies are marked by the error bars
in the upper-right corner (Cayatte \etal\ 1994).  The two DLAs found in groups 
of galaxies are marked in circles.  The size of the points in the left panel 
indicates the intrinsic brightness of the galaxies: $M_{AB}(B) - 5\log h \le 
-19.6$ (large), $-19.6 < M_{AB}(B) - 5\log h \le -18$ (medium), and $M_{AB}(B) 
- 5\log h > -18$ (small).  The size of the points in the right panel indicates 
the redshift interval of the galaxies: $z \le 0.3$ (large), $0.3 < z \le 0.6$ 
(medium), and $z > 0.6$ (small).}
\end{figure}

\newpage

\begin{figure}
\epsscale{1.01}
\plotone{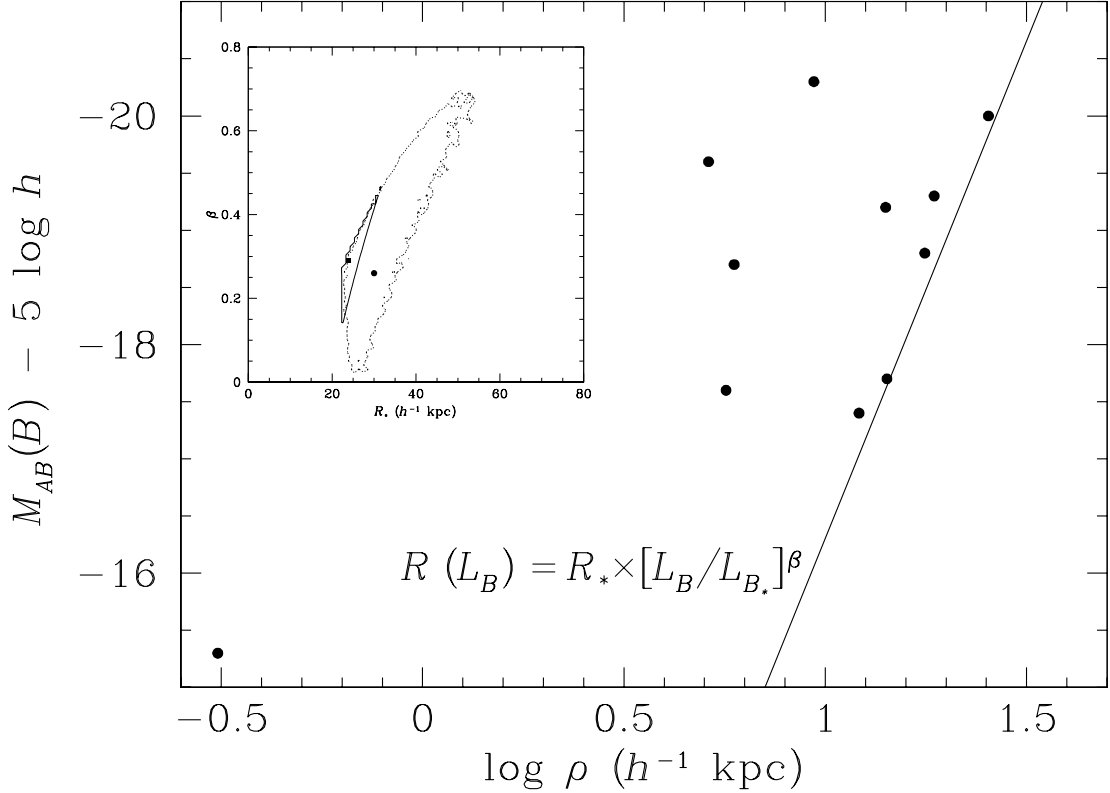}
\caption[]{The distribution of $B$-band absolute magnitude versus impact 
parameter for the 11 galaxies known to produce DLAs.  The data exhibit an
apparent envelope that stretches to larger $\rho$ at brighter $M_{AB}(B)$, 
which may be described by a power-law relation between the \hI\ extent $R$ and
the $B$-band luminosity $L_B$ of the absorbing galaxy.  The solid line shows
the best-fit scaling relation determined for a uniform sphere model using a
maximum likelihood analysis.  The results of the likelihood analysis together
with the corresponding 95\% error contour are indicated by the solid square and
the thick solid line in the inset.  The results of the likelihood analysis for
a uniform disk model is also presented in the inset as indicated by the solid
circle and the thin dotted line.}
\end{figure}

\newpage

\begin{figure}
\plotone{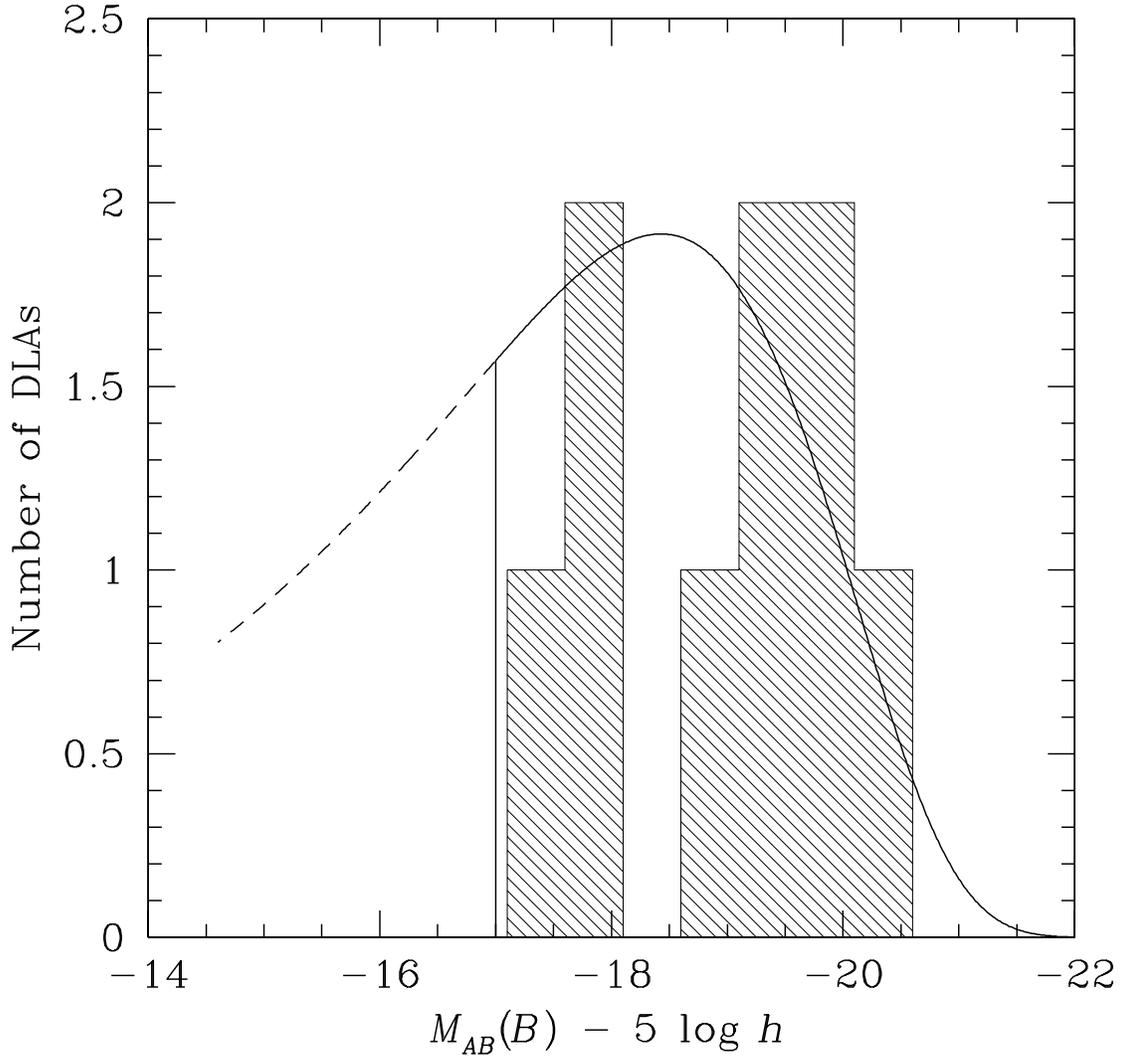}
\caption[]{The incidence of DLAs versus galaxy $B$-band magnitude $M_{AB}(B)$.
The shaded histogram shows the distribution of DLA galaxies from the 
homogeneous sample.  The solid curve shows the predicted incidence of DLAs 
versus galaxy luminosity.  The model was calculated from adopting a Schechter 
luminosity fuctioin, which is characterized by $M_{{AB}_*}(B)=-19.6$ and 
$\alpha=-1.4$, and the best-fit scaling relation determined in using a Monte 
Carlo method as described in \S\ 7.2.  The model has been normalized to match 
the total number of DLA galaxies observed in the homogeneous sample.  The 
dashed curve indicates the expected number of DLAs produced by these fainter 
galaxies if their neutral gas cross section were to be characterized by the 
same scaling relation of the more luminous ones.}
\end{figure}

\newpage

\begin{figure}
\plotone{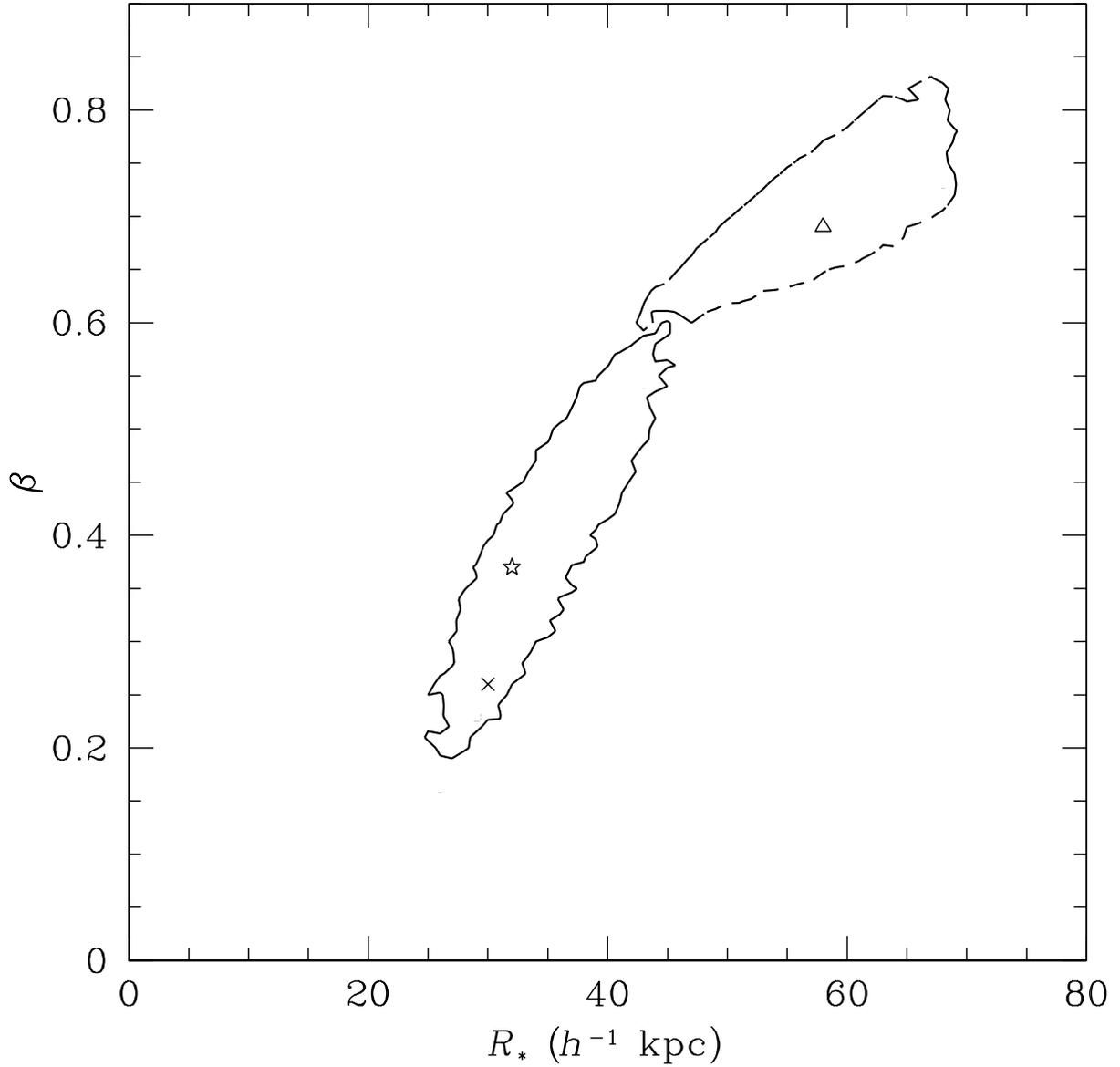}
\caption[]{Results of a Monte Carlo analysis to determine the best-fit scaling
relation between the \hI\ gaseous extent and galaxy $B$-band luminosity for a 
Schechter luminosity function characterized by $M_{{AB}_*}(B)=-19.6$ and 
$\alpha=-1.4$ (Ellis \etal\ 1996).  The triangle represents the scaling 
relation that best describes the data if all galaxies participate in 
absorption.  The dashed line indicates the corresponding 95\% error contour.  
The star represents the scaling relation that best describes the data if we 
impose a cut-off in the gas cross section for galaxies fainter than 
$M_{AB}(B)-5\,\log h=-17$.  The solid line indicates the corresponding 95\% 
error contour.  For comparison, the cross indicates the best-fit scaling 
relation determined in Figure 13 for a disk model using the inhomogeneous DLA 
sample.}
\end{figure}

\newpage

\begin{figure}
\epsscale{1.01}
\plotone{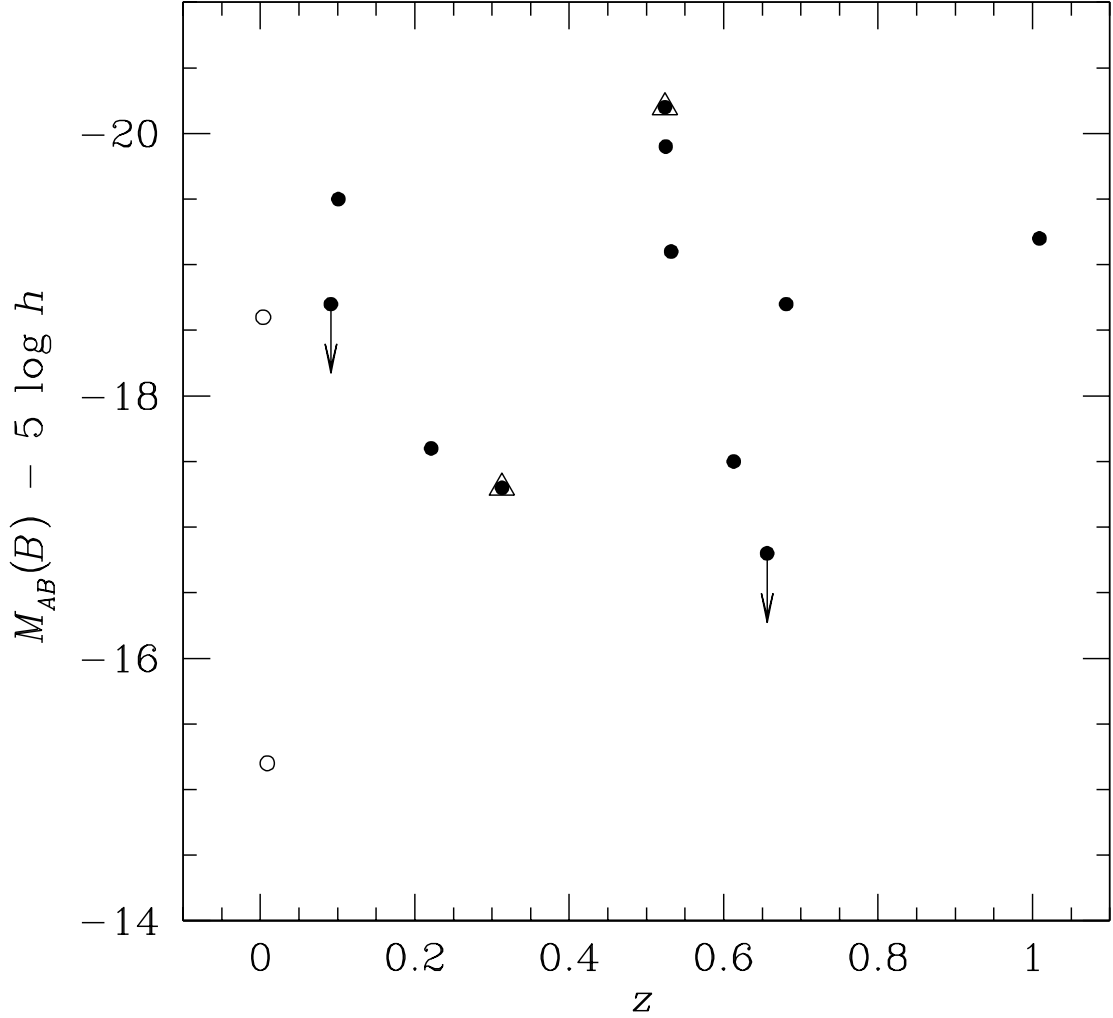}
\caption[]{The distribution of rest-frame $B$-band absolute magnitude 
versus redshift for 13 known DLAs.  Points with arrows are the DLAs for which 
the absorbing galaxies have not been found after extensive searches.  The 
magnitudes indicated by the points are the brightness limits of the underlying
DLA galaxies as set by the depths of the existing surveys (see the text). 
Solid points indicate those DLAs in the homogeneous sample and open circles
indicate the two systems found in targeted DLA surveys.  The two systems found
to be associated with groups of galaxies are marked with open triangles.}
\end{figure}

\end{document}